\documentclass[fleqn,usenatbib]{aa}
\usepackage[T1]{fontenc}

\DeclareRobustCommand{\VAN}[3]{#2}
\let\VANthebibliography\thebibliography
\def\thebibliography{\DeclareRobustCommand{\VAN}[3]{##3}\VANthebibliography}

\usepackage{graphicx}	
\usepackage[export]{adjustbox}
\usepackage{bm}	

\usepackage{amsmath}	
\usepackage{hyperref}
\usepackage{comment}

\usepackage{color, xcolor}

\newcommand{\ik}[1]{{\color{black} #1}}

\def\CL{\textcolor{black}{SRGe~CL0512.7+3712}}

\usepackage{txfonts}
\begin{document} 

   \title{New galaxy cluster in the Zone of Avoidance {SRGe~CL0512.7+3712}. Discovery and multi-wavelength characterization
   }
    \titlerunning{SRGe~CL0512.7+3712}
    \authorrunning{}

   \author{Ildar~Khabibullin \inst{\ref{in:MPA},\ref{in:RPC},\ref{in:IKI}}
\and 
Eugene~Churazov \inst{\ref{in:MPA},\ref{in:IKI}}
\and
   Natalya~Lyskova
\inst{\ref{in:IKI}}     
    \and 
    Ilfan~Bikmaev
\inst{\ref{in:KFU},\ref{in:TAS}}
    \and
    Eldar~Irtuganov
\inst{\ref{in:KFU},\ref{in:TAS}}
    \and
    Mikhail~Suslikov
\inst{\ref{in:KFU},\ref{in:TAS}}
    \and
Igor~Zaznobin\inst{\ref{in:IKI}}
    \and 
    William~R.~Forman\inst{\ref{in:CfA}}
    \and
    Ralph~Kraft\inst{\ref{in:CfA}}
    \and
    Rashid~Sunyaev\inst{\ref{in:MPA},\ref{in:IKI}}
    \and
    Alexei Moiseev\inst{\ref{in:SAO},\ref{in:IKI}}
    \and
    Arkadiy Sarkisyan\inst{\ref{in:SAO}}
          }
   \institute{
   Max Planck Institute for Astrophysics, Karl-Schwarzschild-Str. 1, D-85741 Garching, Germany \label{in:MPA}
\and
Rudolf Peierls Centre for Theoretical Physics, Department of Physics, University of Oxford, Clarendon Laboratory, Parks Rd, Oxford, OX1 3PU, United Kingdom \label{in:RPC}
\and
Space Research Institute (IKI), Profsoyuznaya 84/32, Moscow 117997, Russia\label{in:IKI}
   \and
Center for Astrophysics, Harvard \& Smithsonian, 60 Garden St, Cambridge, MA 02138, USA \label{in:CfA}
\and
Kazan Federal University,  Kremlevskaya Str. 18,  420008 Kazan, Russia \label{in:KFU}
\and
Tatarstan Academy of Sciences, Bauman Str. 20, 420111, Kazan, Russia \label{in:TAS}
\and
Special Astrophysical Observatory, Russian Academy of Sciences, Nizhny Arkhyz 369167, Russia\label{in:SAO}   
             }

   \date{\today}

\abstract{
~~~~The census of massive clusters of galaxies in the local Universe is almost complete, thanks to their prominent observational signatures at X-ray, optical, and sub-mm wavelengths. Nevertheless, a number of such systems \ik{are} likely to be missing and hidden behind the plane of our Galaxy, where high interstellar absorption as well as strong contamination by foreground stellar and diffuse sources prevent detection of even the brightest and the most massive ones. Here we report the discovery and multiwavelength characterization of such a cluster in the zone of avoidance (ZoA) \CL~ in the data of SRG/eROSITA all-sky survey. Combining the data of radio, optical, and infrared surveys, we identify overdensity of possible red sequence galaxies, as well as the candidate brightest cluster galaxy. Follow-up optical and X-ray observations confirm that the newly found object is a massive ($M_{500c}=(4-5)\cdot 10^{14}M_{\odot}$, $kT\approx 5 $ keV) galaxy cluster at redshift $z=0.0745$ with possible indications of unrelaxed dynamical scale. Location and elongation of this cluster is consistent with an expectation from the large-scale structure at this redshift, and it might be a part of an extended overdensity of such objects in the Galactic Anticenter direction. Examination of X-ray, radio, and infrared data in the locations of ZoA, where similar objects are expected to be found based on the {large-scale structure} properties, might reveal another $\sim10$ clusters at this redshift in future.}
\maketitle
%

%
\section{Introduction}
\label{s:intro}

~~~Massive galaxy clusters, containing hundreds of Milky Way-like galaxies, mark strongly overdense knots of the Cosmic Web arising at the intersection of large-scale ($\gtrsim$10 Mpc) structures like sheets and filaments \citep[e.g.,][]{2012ARA&A..50..353K}. Some of the most extreme and well-studied examples are Coma and Perseus clusters and their direct vicinities \citep[e.g., ][]{2021A&A...656A.144B}, as well as the Virgo cluster, the nearest cluster having direct relation to our Galaxy's (and the Local Group's) large-scale environment \citep[e.g.,][]{2019MNRAS.486.3951S}. 

Great progress in characterization of the galaxy cluster population in the local Universe ($z\lesssim0.1$) has been achieved thanks to sensitive large-area surveys in the optical \citep[e.g.,][]{1989ApJS...70....1A,2009ApJS..183..197W,2014ApJ...785..104R,2024ApJS..272...39W}, infrared \citep[IR, ][]{2015AJ....149..171T}, X-ray \citep[][]{1998MNRAS.301..881E,2001A&A...369..826B,2004A&A...425..367B,2023MNRAS.526.3757K,2024A&A...685A.106B,2024A&A...688A.187S} and microwave \citep[][]{2016A&A...594A..27P,2019A&A...626A...7T,2024MNRAS.531.1998V} spectral bands. However, a large part of the sky is strongly affected by extinction in the Galactic plane, as well as by the confusion of numerous point-like and extended sources located there \citep[e.g., ][]{2002ApJ...580..774E}. A region of Galactic latitudes $|b|<20^\circ$ is commonly referred to as the Zone of Avoidance (ZoA), where only the brightest (most massive and nearby) objects have been detected and identified, collectively dubbed as ``Clusters In the ZoA'' \citep[CIZA, ][]{2002ApJ...580..774E,2007ApJ...662..224K}. Improvements in the completeness of our knowledge of the CIZA population would help better constrain and reconstruct our local environment \citep[e.g., ][]{2025A&A...695A..59B}, but this task remains a challenge so far.

Some of the ZoA clusters turned out to possess prominent sub-structures when observed in radio, with the Sausage \citep[CIZA~J2242.8+5301, ][]{2007ApJ...662..224K,2010Sci...330..347V} and 3C~129 \citep[CIZA~J0450.0+4501, ][]{2002ApJ...580..774E,2020A&A...644A.107R} clusters being \ik{among the best examples} examples. Although this is not unexpected for the massive clusters and groups \citep[e.g., ][ for a review]{2019SSRv..215...16V}, it also highlights a great synergy of X-ray and radio observations for the identification and exploration of these \ik{rare} objects.

Here we report a discovery, redshift measurement, and follow-up characterization of a new nearby (at redshift $z=0.0745$) massive galaxy cluster \CL~ in the ZoA enabled by the data of SRG/eROSITA all-sky survey \citep[][]{2021A&A...647A...1P,2021A&A...656A.132S}. Using archival radio and infrared images, we identified the brightest cluster galaxy and measured its redshift through dedicated spectroscopic observations, while follow-up observations with the Chandra observatory allowed us to obtain estimates for the mass of this cluster based on luminosity-mass and temperature-mass relations. These findings place the newly found object among the most massive clusters \ik{in the ZoA} known so far.

The paper is structured as follows: we describe the discovery and identification of the possible BCG in Sect.~\ref{s:discovery}. We present results of follow-up optical and X-ray observations in Sect.~\ref{s:followup}. We discuss these findings in Sect.~\ref{s:discussion} and summarize our conclusions in Sect.\ref{s:conclusion}. 
Throughout the paper, we adopt a $\Lambda$ cold dark matter cosmology with $\Omega_{\rm M} = 0.3,~\Omega_{\rm \Lambda} =0.7$ and $H_0=70$ km s$^{-1}$ Mpc$^{-1}$. 
The cluster's radius $R_{500c}$ is defined as the radius within which the enclosed matter overdensity exceeds the critical density of the Universe $\rho_{c}(z)=\frac{3H^2(z)}{8\pi G}$ at redshift $z$ by a factor of 500, and the corresponding mass within this radius is $M_{500c}=(4\pi/3)\cdot 500~\rho_c \cdot R_{500c}^3\approx 3\cdot 10^{14}M_{\odot}\left(\frac{R_{500c}}{\rm 1~Mpc}\right)^{3}(1+z)^3$.
\section{Discovery of \CL}
\label{s:discovery}
\begin{figure}
    \centering
    \includegraphics[angle=0,,clip=true,trim=2.1cm 7cm 2.1cm 4.5cm,width=0.9\columnwidth]{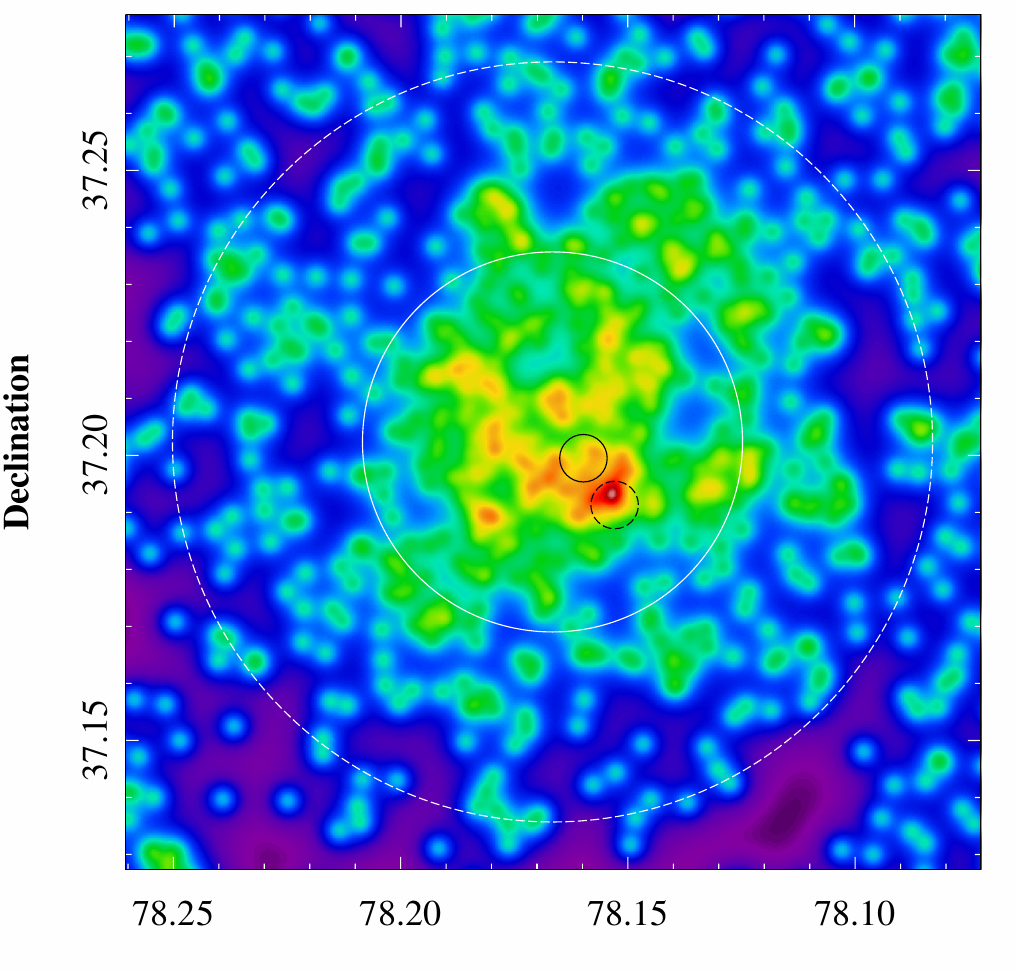}
    \includegraphics[angle=0,,clip=true,trim=2.1cm 7cm 2.1cm 4.5cm,width=0.9\columnwidth]{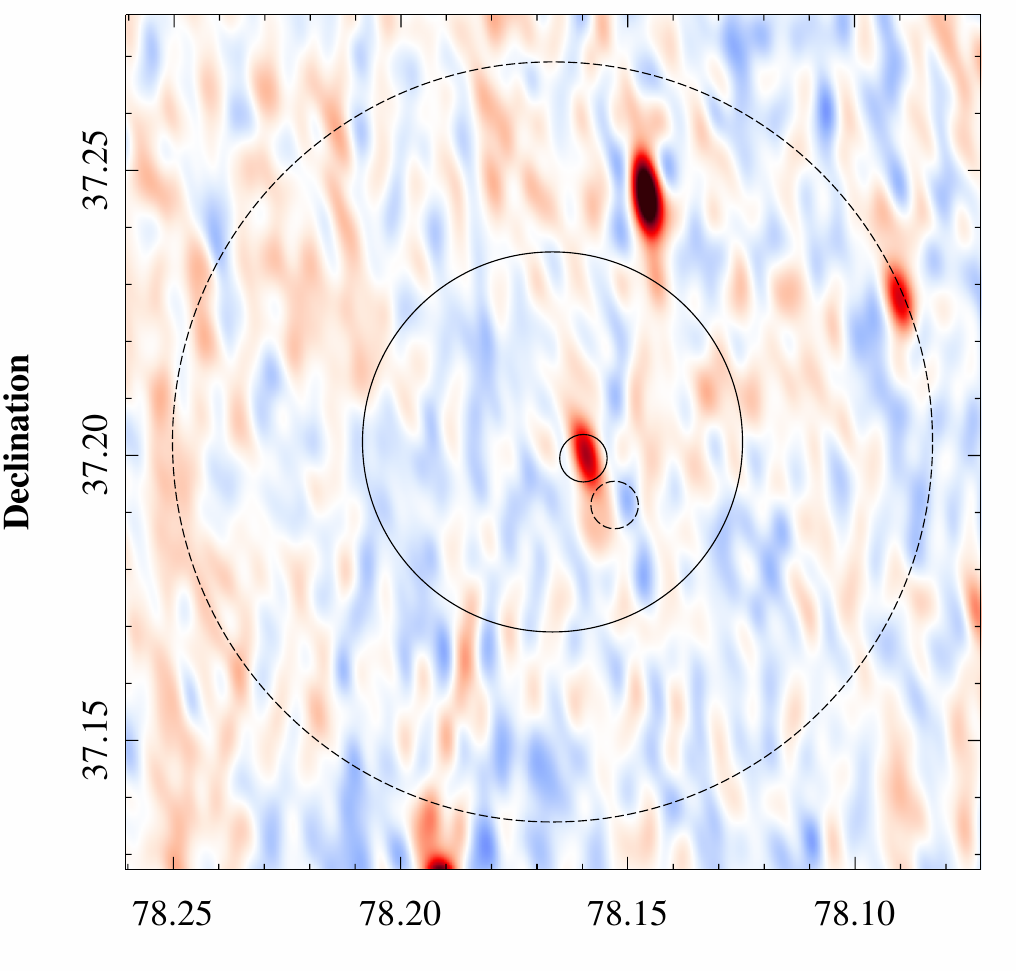}
    \includegraphics[angle=0,,clip=true,trim=2.1cm 6cm 2.1cm 
    4.5cm,width=0.9\columnwidth]{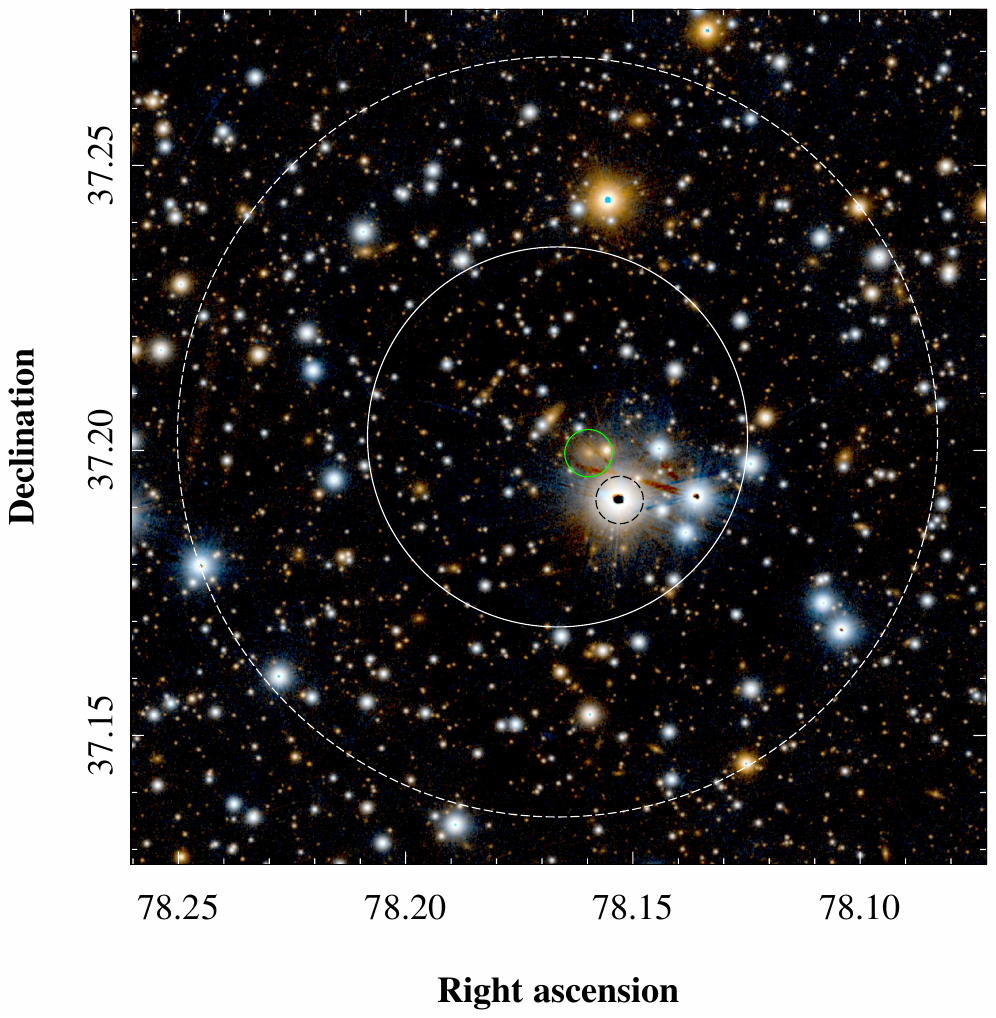}   
    \caption{X-ray (SRG/eROSITA, 0.4-2.3 keV, top), radio (RACS-low, middle), and optical (Pan-STARRS1, zrg, bottom) images of \CL. The big circles mark radii equal to 2$'$ (solid) and 4$'$ (dashed). The small circles ($15''$ radius) mark locations of the candidate BCG 2MASS~J05123831+3711581 (solid) and the bright foreground star HD~280580 (dashed).
    }
    \label{fig:srge_ps1}
\end{figure}

In this section, we describe the data that enabled the discovery of \CL. Essentially, these data are provided by large area X-ray, radio, and optical/IR surveys of the Galactic plane, and the presented scheme might be used as a primer for future searches for similar objects.

\subsection{X-ray emission in SRG/eROSITA All-Sky Survey}
The all-sky survey conducted by the eROSITA telescope \citep[][]{2021A&A...647A...1P} onboard SRG observatory \citep[][]{2021A&A...656A.132S} delivered sensitive (with the typical point source detection limit of $2\cdot10^{-14}$ erg s$^{-1}$ cm$^{-2}$ in 0.4-2.3 keV band ) and 
complete coverage of the whole sky, including the ZoA region, commonly defined as a broad stripe at Galactic latitudes $|b|\leq20^{\circ}$. A systematic search for very extended sources (at least several arcmin in size) in this region, aimed at finding objects likely associated with supernova remnants (SNRs), has been performed after modeling and subtraction of the detected compact sources and smoothing with a Gaussian window of an arcmin size \citep[e.g.,][]{2023MNRAS.521.5536K}.  In addition to finding new SNR candidates \citep[e.g. G121.1-1.9,][]{2023MNRAS.521.5536K}, the program resulted in the discovery of an extended source (at least 6$'$ in diameter) at Galactic coordinates $(l,b)=(169.30^\circ,-01.18^\circ)$ (see Fig.~\ref{fig:srge_ps1}) with no known SNR being located at this position \citep[e.g.,][]{2025JApA...46...14G}. Significant emission was detected above 2.3 keV, with the overall spectral shape of the emission extracted from the central 4$'$ being consistent with thermal emission with a temperature above 5 keV (see Sect.~\ref{ss:chandra}).  

\subsection{Radio emission inside X-ray region}

No matching diffuse radio emission has been found in the archival images of Canadian Galactic Plane Survey \citep[CGPS][]{2003AJ....125.3145T}, Rapid ASKAP Continuum Survey (RACS) by Australian Square Kilometre Array Pathfinder (ASKAP) at 887.5 MHz \citep[RACS-low,][]{2020PASA...37...48M} and 1367.5 MHz \citep[RACS-mid,][]{2023PASA...40...34D}, and LOFAR Two-meter Sky Survey (LoTSS) DR3 at 150 MHz \citep[][]{2026arXiv260215949S}, although these surveys have proven to be very efficient in detecting faint diffuse radio emission at scales of several arcmin \citep[e.g.,][]{2026PASA...43...12K}. Instead, an apparently compact radio source is located close to the center of X-ray emission (see Fig.~\ref{fig:srge_ps1}), which, however, is not included in the RACS-mid catalog \citep[][]{2024PASA...41....3D}, meaning that it is fainter than 2 mJy at 1367.5 MHz (95\% completeness limit of the RACS-mid survey). This limit is also consistent with the fluxes of the faintest nearby sources present in the catalog (e.g., RACS-MID~J051212.0+371555 to the northwest with the flux density $F_{1367}=2.3\pm0.53$~mJy at 1367.5 MHz). 

At the same time, this source is detected with high significance at lower frequencies in the data of LoTSS DR3 \citep[][]{2026arXiv260215949S} as ILTJ051238.28+371158.5 with an integrated flux density $F_{144}=4.6\pm0.1$~mJy at 144~MHz. It is also classified as a multi-gaussian source with spatial extent $FWHM\approx(6.5-7.7'')$\citep[][]{2026arXiv260215949S}. Assuming that the flux density of this source is just below the sensitivity threshold, e.g. $F_{1367}\sim1$~mJy at 1367.5 MHz, the estimated spectral index $\alpha\approx0.67$ between 144 and 1367.5 MHz assuming $F_\nu\propto\nu^{-\alpha}$. New releases of the RACS catalog at lower frequencies (887.5 MHz)  should help better constrain the slope \citep[currently available RACS-low catalog, ][does not cover the region of interest here]{2021PASA...38...58H}.

\subsection{Candidate brightest central galaxy}

Examination of the optical images from the Pan-STARRS1 survey \citep[][]{2016arXiv161205560C} shown in Fig.~\ref{fig:srge_ps1} reveals that the compact radio source coincides with an early-type galaxy 2MASS~05123831+3711581 \citep[][]{2006AJ....131.1163S}, while a number of smaller galaxies are observed in the region of X-ray emission (see Fig.~\ref{fig:srge_ps1}). Also, a flaring star HD~280580 \citep[an oscillating red giant with $V\approx 9 $, ][]{2021ApJ...919..131H} is projected very close to its center, and it might be responsible for a certain fraction of observed X-ray emission (especially the softer portion of it). Since this region is characterized by the large integrated column density of absorbing gas, equivalent to the reddening $E(B-V)=1.1$ or total hydrogen column density $N_{\rm H}=0.75\cdot10^{22}$~cm$^{-2}$\citep[][]{2013MNRAS.431..394W}, optical emission of the galaxies in this direction is strongly attenuated and is quite faint. Still, another apparently extended disk-like galaxy, 2MASX~05124037+3712199 \citep[][]{2000AJ....119.2498J} is present close to the center of X-ray emission, in addition to a multitude of smaller galaxies \ik{located} around. 

\begin{figure}
    \centering
    \includegraphics[angle=0,,clip=true,trim=2.1cm 6.1cm 2.1cm 4.5cm,width=0.99\columnwidth]{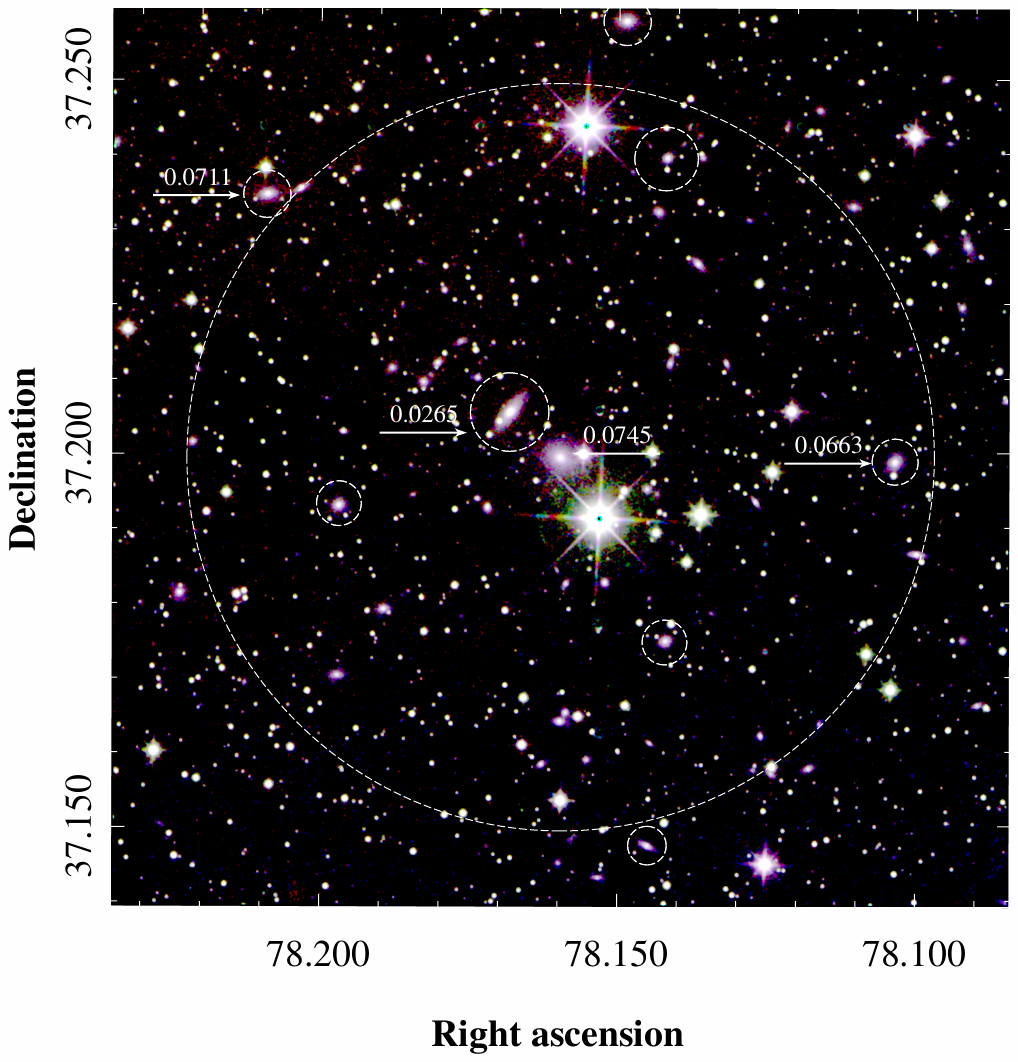}   
    \caption{Composite NIR image based on UKIDSS DR11 GPS \citep[][]{2008MNRAS.391..136L} data in $JHK$ bands. Extended sources from 2MASS survey \citep{2000AJ....119.2498J} are marked as circles with radius being 3$r_{3\sigma}$, where $r_{3\sigma}$ is the semi-major axis of the $K$-band isophotal $3\sigma$ ellipse. The large circle is centered on  2MASS~05123831+3711581 and has a radius of $3'$. Note that this galaxy is not included in the 2MASX catalog, probably because of contamination from the bright nearby star HD~280580. \ik{Arrows mark galaxies for which optical spectra were obtained with the measured redshift values given (see Sec.~\ref{ss:opticalspectra}).}
    }
    \label{fig:ukidss}
\end{figure}

The attenuation problem is strongly alleviated by switching to the near infrared (NIR) band, in particular using data from the 2MASS \citep[][]{2006AJ....131.1163S} and UKIDSS \citep[][]{2007MNRAS.379.1599L} surveys. Figure~\ref{fig:ukidss} shows a composite NIR image in $JHK$ bands from the UKIDSS Galactic Plane Survey \citep[][]{2008MNRAS.391..136L} with an overlay of extended sources (larger than a few arcsec in size) present in the 2MASS Extended Source Catalog \citep[2MASX,][]{2000AJ....119.2498J}. Note that 2MASS~05123831 + 3711581 galaxy is not present in this catalog, possibly due to contamination from the HD~280580 star and modest PSF size of 2MASS ($\sim2''$). The imaging quality of the UKIDSS GPS data is much better ($\lesssim0.8''$), and one can detect many more extended sources (see Fig.~\ref{fig:ukidss}).   

\subsection{Overdensity of extended infrared sources in 2MASX}
\label{ss:ukidss}

Given that the 2MASX catalog includes only the brightest and most extended sources, and we can see that at least some candidate member galaxies are present in it (see Fig.~\ref{fig:ukidss},), one can check if there is a noticeable overdensity of similar sources coincident with the extended X-ray emission. In particular, all 2MASX sources shown in Fig.~\ref{fig:ukidss}, except for the apparently disk-like 2MASX~05124037+3712199, have $r_{3\sigma}<5''$, so we can define "candidate" member galaxies as all 2MASX sources with $r_{3\sigma}<5''$. 

\begin{figure}
    \centering
    \includegraphics[angle=0,,clip=true,trim=2.5cm 6.5cm 2.2cm 4.6cm,width=0.99\columnwidth]{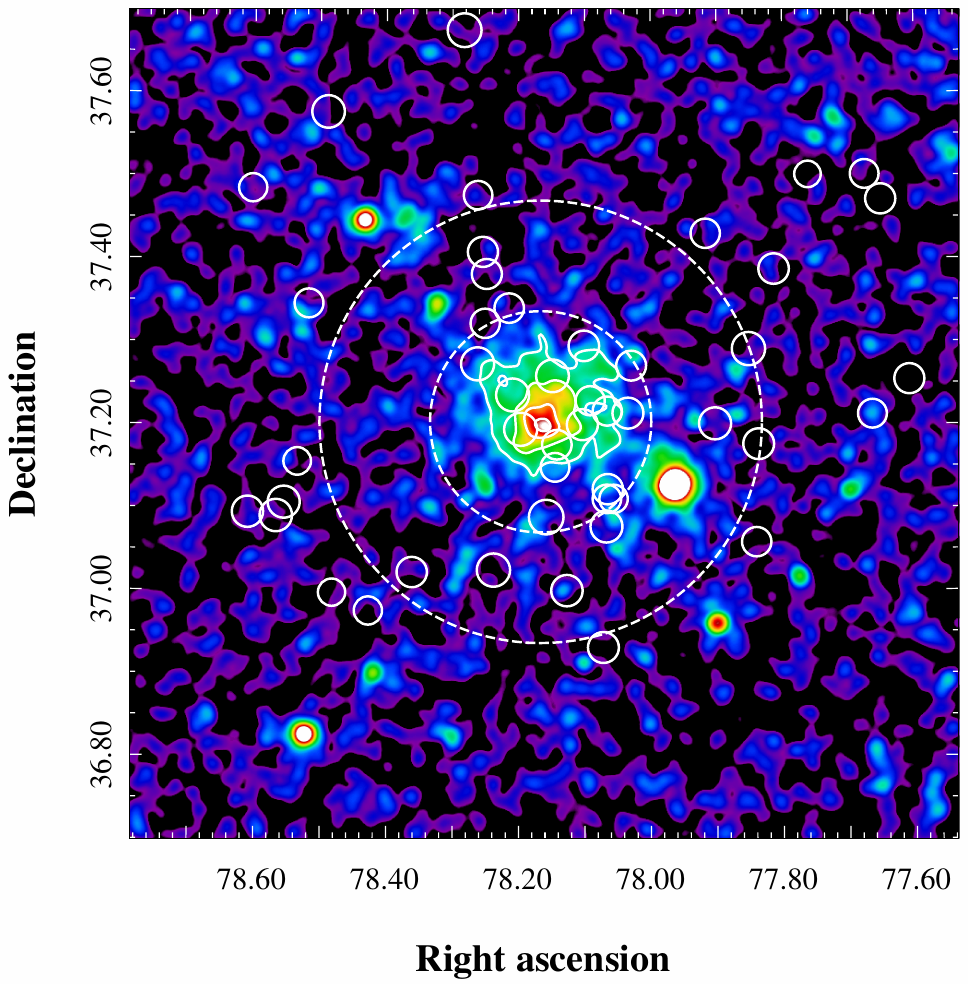}   
    \caption{Large scale ($1^\circ\times1^\circ$) SRG/eROSITA 0.4-2.3 keV (particle background-subtracted and exposure-corrected) image with an overlay of all sources from the 2MASX catalog with $r_{3\sigma}<5''$. The X-ray image is on a sqrt scale and was smoothed with a Gaussian window with $\sigma=24''$. The dashed circles are $8'$ and $16'$ in radius. \ik{The white contours show surface brightness levels 5, 10, 20, and 30 times the background level estimated at the boundaries of this image.}
    }
    \label{fig:srge_2MASX}
\end{figure}

Figure~\ref{fig:srge_2MASX} shows a similar to Fig.~\ref{fig:srge_ps1} smoothed large scale soft X-ray image from SRG/eROSITA with an overlay of 2MASX sources with $r_{3\sigma}<5''$. An apparent overdensity of such sources is obvious not only in the center of diffuse X-ray emission, e.g., within 8', but also even on a twice larger scale, as indicated by dashed concentric circular regions in Fig.~\ref{fig:srge_2MASX}. 

Since the UKIDSS catalog is much deeper and has better spatial resolution than 2MASX, a similar procedure performed on the full UKIDSS catalog of extended sources would result in a picture completely dominated by fainter sources, many of which will likely be background galaxies. In order to proceed, one needs to define proper selection criteria, for instance those that take into account photometric properties as well. The most common approach uses identification of the so-called red sequence of the cluster galaxies on a color-magnitude plane based on optical or NIR photometry \citep[e.g., ][ and references therein]{2020A&A...644A.107R}.  

\subsection{Identification of optical and NIR red sequences}
\label{ss:reds}

\begin{figure}
    \centering
        \includegraphics[clip=true,trim=0cm 0cm 0cm 0cm,width=0.8\columnwidth]{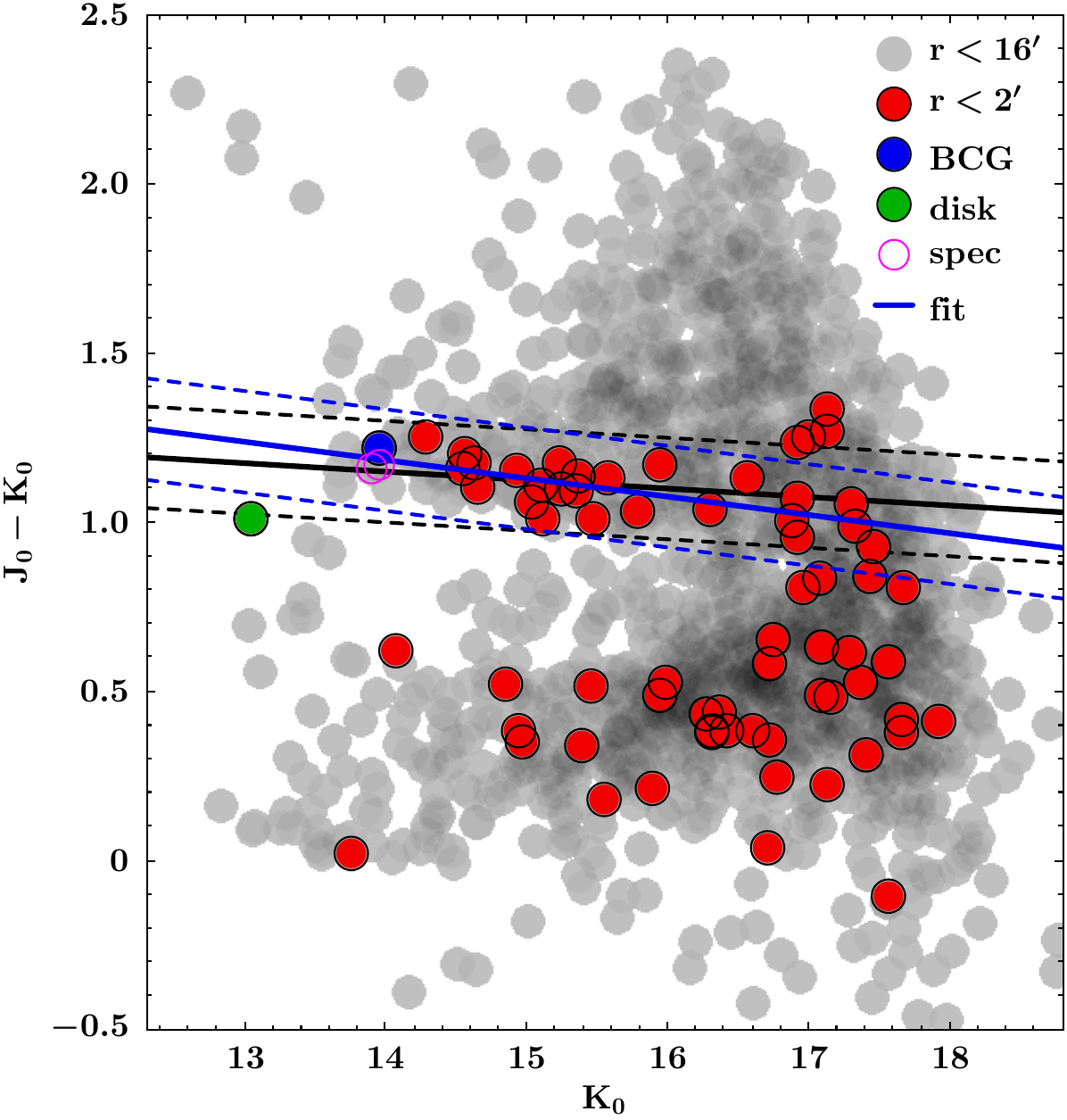}
        \includegraphics[clip=true,trim=-4cm -2cm 0cm 0cm,width=0.9\columnwidth]{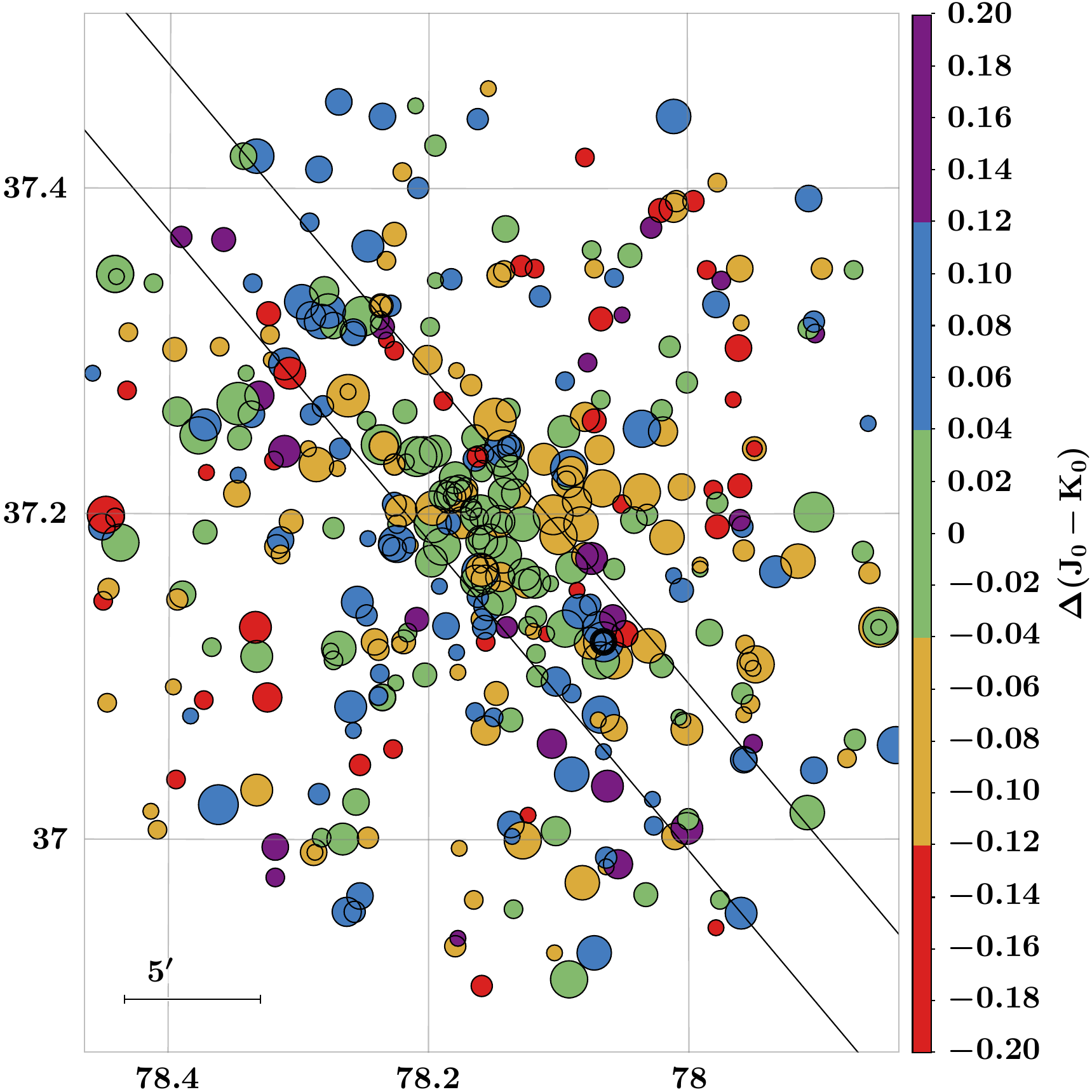}
        \begin{picture}(0,0)
\put(-140,0){\rotatebox{0}{\large \textcolor{black}{Right Ascension}}}
\put(-220,90){\rotatebox{90}{\large \textcolor{black}{Declination}}}
\end{picture}
    \caption{Identification of the candidate red sequence of galaxies based on UKIDSS~GPS. \textit{Top.} NIR magnitude-color for bona fide non-star sources in UKIDSS~GPS catalog within $2'$ (red) and $16'$ (gray) of the center of X-ray emission. The candidate BCG galaxy is shown in blue, while the disk galaxy is shown in green, \ik{and two additional galaxies selected for spectroscopy (see Sec.~\ref{ss:opticalspectra}) are shown in magenta.} The solid black line shows the red-sequence relation in the form $J_0-K_0=-0.025K_0+1.5$, while the dashed lines mark a $\Delta (J_0-K_0)=0.15$ interval from it. Similarly, the solid blue line shows the best-fit red-sequence relation in the form $J_0-K_0=-0.054K_0+1.94$ with the dashed lines marking a $\Delta (J_0-K_0)=0.15$ interval from it. All galaxies between the dashed blue lines are considered as red-sequence candidates. \textit{Bottom.} Spatial distribution of the red-sequence candidate galaxies within $16'$ with their size $R=50-2.5K_0$ arcsec and color-coding according to the $J_0-K_0$ color difference from the best-fit red-sequence relation. Black lines mark direction with the position angle $PA=40^\circ$ along which the distribution of the candidate galaxies appears to be elongated.}

    \label{fig:ukidss_redseq}
\end{figure}
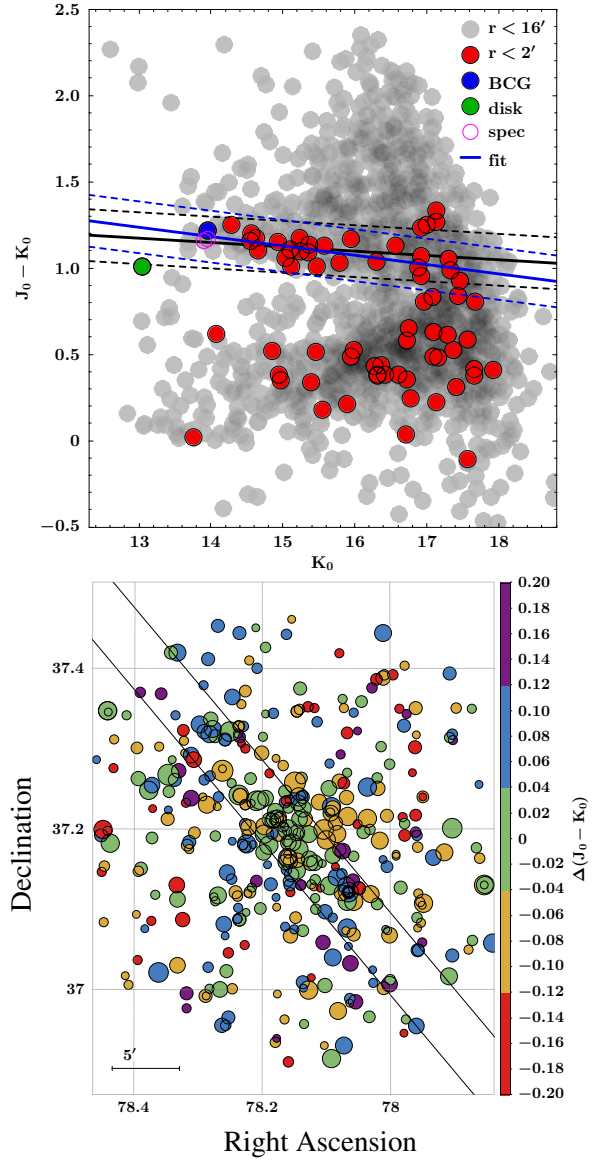

Following previous studies on galaxy clusters \ik{in the ZoA} (e.g. the 3C~129 cluster), as well as rich extraplanar clusters like Coma \citep[][and references therein]{2020A&A...644A.107R}, we identify an NIR red sequence based on absorption-corrected $J_0$ and $K_0$ magnitudes from the UKIDSS GPS catalog. The absorption-corrected magnitudes are computed based on the observed magnitudes using $E(B-V)=1.1$ and standard coefficients relating it to $A_K$ and $A_J$ \citep[
][]{2020A&A...644A.107R}.  We select only sources with listed probability being a star ($p^*<0.005$) to avoid contamination by stellar sources \citep[][]{2008MNRAS.391..136L}. Starting by considering only sources within $2'$ from the center of the X-ray emission, we can clearly see a sequence of galaxies following the expected behavior with the candidate BCG galaxy at the brightest end of it (red points in Fig.~\ref{fig:ukidss_redseq}). 

The black line in Fig.~\ref{fig:ukidss_redseq} shows a simple linear relation similar to the one found for the 3C~129 and Coma clusters in the form of $J_0-K_0=-0.025K_0+1.5$ \citep[][]{2020A&A...644A.107R}, with all bright members ($K_0$<16.6) of this sequence being within $\Delta(J_0-K_0)=0.15$  around this relation (as shown by the black \ik{dashed} lines in Fig.~\ref{fig:ukidss_redseq}). If, instead, the slope and the offset of this relation are allowed to be fitted, the relation takes the form $J_0-K_0=-0.053K_0+1.94$ (blue solid line in Fig.~\ref{fig:ukidss_redseq}) with the rms deviation of 0.05 (based on 20 brightest sources). Although these relations are slightly different from each other, the resulting selections of the red-sequence galaxies are almost identical. It is noteworthy that the disk galaxy 2MASX~05124037 + 3712199 (shown in green in Fig.~\ref{fig:ukidss_redseq}) is located significantly off this relation, and no similar sequence of fainter galaxies might be associated with it.

Also, the spatial distribution of the candidate red sequence galaxies shows elongation along the south-west to north-east direction, as already hinted in Fig.~\ref{fig:srge_2MASX}. We highlight this direction defined by the position angle $PA=40^\circ$ (defined in the standard counter-clockwise manner from the celestial north direction).  Interestingly, the inner part of the X-ray emission shows signatures of elongation in the direction perpendicular to that, which might reflect a more recent assembly history of the cluster \citep[see an example of the Perseus cluster, ][]{2026A&A...707A.381C}.

\begin{figure}
    \centering
    \includegraphics[angle=0,,clip=true,trim=2cm 6.8cm 2.1cm 4.cm,width=0.99\columnwidth]{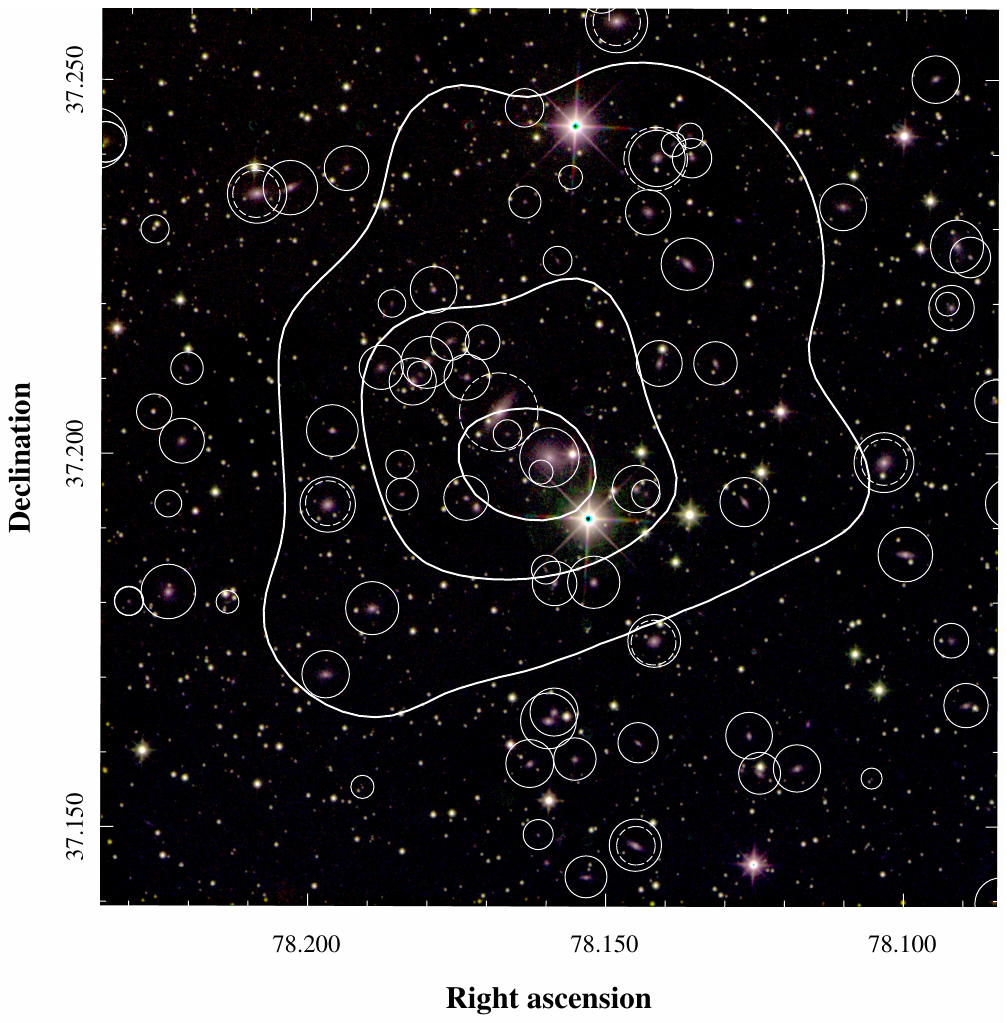}  
    \caption{Infrared picture of the central part of \CL. Composite NIR image based on UKIDSS DR11 GPS data with an overlay of contours of X-ray surface brightness from eROSITA (0.4-2.3 keV) and candidate red-sequence galaxies (solid white circles with radius $R=50-2.5K_1$ arcsec). Dashed circles show the same extended sources from the 2MASX catalog as in Fig.~\ref{fig:ukidss}, confirming that all of them, except for the disk galaxy in the center, are likely cluster members and justifying the usage of similar galaxies as tracers for the environment of the cluster. 
    }
    \label{fig:ukidss_srgecont_2masx}
\end{figure}

\section{Follow-up observations and physical characterization}
\label{s:followup}
%

Morphological appearance, red-sequence determination, and detection of a radio source coincident with 2MASS~05123831+3711581 strengthen the case for it being a potential BCG, given that supermassive black holes in BCGs of massive galaxy clusters are often able to accrete at a level sufficient for powering radio activity. Spectroscopic optical observations, although being rather challenging because of high attenuation and density of bright foreground stellar sources, are vital for the determination of the cluster's redshift and spectroscopic confirmation for the candidate member galaxies. 

\subsection{Optical spectroscopy of the candidate BCG}
\label{ss:opticalspectra}
\ik{
An optical red sequence, which is particularly useful to identify the best targets for the spectroscopic observations needed to determine the redshift of the candidate cluster, can be built in a similar manner based on the Pan-STARRS1 DR2 photometric data \citep[][]{2020ApJS..251....6M}.}

\ik{
By using the Pan-STARRS1 map (shown in Fig.~\ref{fig:srge_ps1}), we have selected about 20 galaxies in the $6'\times6'$ field centered at BCG candidate 2MASS~05123831+3711581, and extracted their griz Kron magnitudes from the Pan-STARRS DR2 \citep[][]{2020ApJS..251....6M}. Observed griz magnitudes were corrected for extinction $E(B-V) = 1.1$ in the direction of \CL. We constructed a red sequence $r_0-(g-r)_0$ and selected several galaxies with $(g-r)_0=1\pm0.1$ mag as cluster members for spectroscopy. Spectroscopic observations of the candidate brightest cluster galaxy of \CL~  were carried out with (1) the Russian-Turkish 1.5-m telescope (RTT-150) at the {Turkish National Observatories, Observatory in Antalya, Turkey}, and (2) the 6-m BTA telescope of the {Special Astrophysical Observatory of the Russian Academy of Sciences} (see Appendix~\ref{s:optical} for details).}

Redshifts of four galaxies have been measured (with 0.001 accuracy), showing that among the two apparently brightest galaxies, 2MASX~J05124037+3712199 galaxy is located at $z=0.0265$, while the candidate BCG 2MASS~J05123831+3711581 is at $z=0.0745$. The other two fainter galaxies (see Fig.~\ref{fig:ukidss}), 2MASS~J05122483+3711555 and 2MASS~J05125014+3714048, are at $z=0.0663$ and $z=0.0711$, respectively. As a result, we confirm that 2MASX~05124037+3712199 is a foreground galaxy, and redshifts of three other galaxies are close to each other and consistent within rms $\sigma\approx1000$ km~s$^{-1}$. 

At this distance, 1 arcmin corresponds to 86 kpc \citep[assuming standard flat $Planck$ cosmology, ][]{2020A&A...641A...6P}, meaning that 1 Mpc, which is close to the characteristic $R_{500c}$ radius for massive ($M_{500c}>2\times10^{14}M_{\odot}$) clusters, corresponds to 12'. Consistent with the average picture of massive galaxy clusters \citep[][]{2023MNRAS.525..898L}, diffuse X-ray emission of \CL~ is clearly visible against the background up to approximately this radius (see Fig.~\ref{fig:srge_2MASX}), while a noticeable overdensity of galaxies can be traced to significantly \ik{larger} radii.

\subsection{Chandra characterization of \CL}
\label{ss:chandra}
\begin{figure}
    \centering
    \includegraphics[angle=0,,clip=true,trim=2.3cm 7.5cm 2.9cm 6cm,width=0.9\columnwidth]{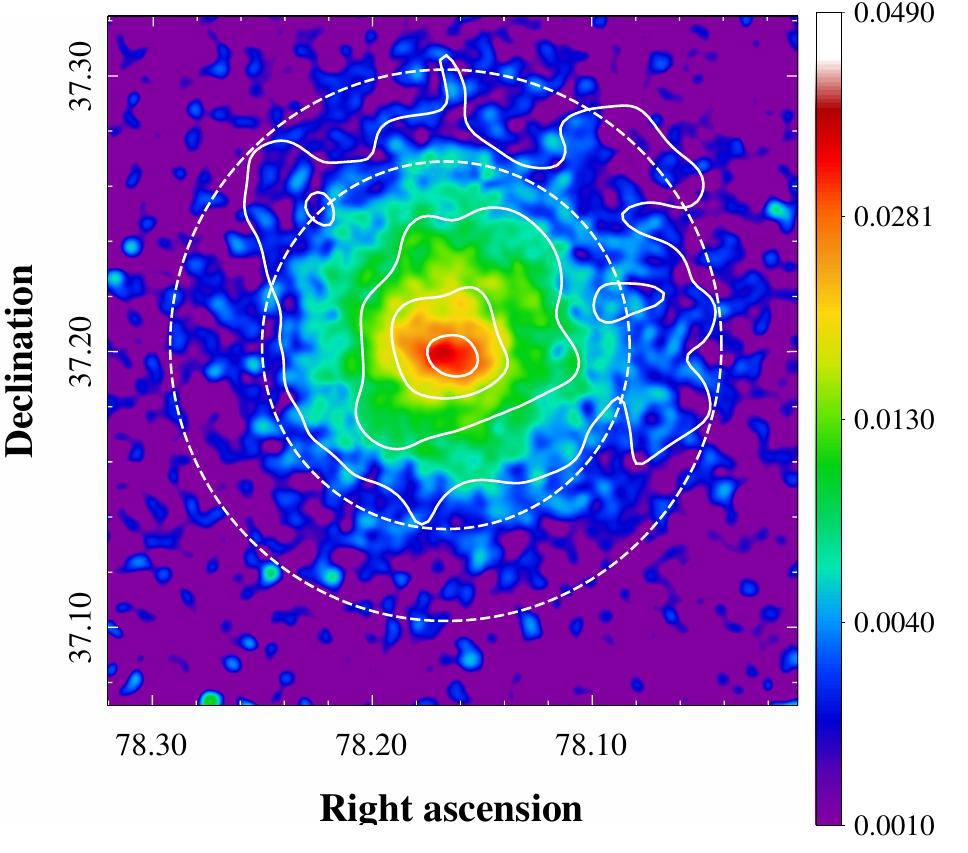}
\includegraphics[angle=0,,clip=true,trim=1.3cm 5.cm -2cm 3.5cm,width=0.9\columnwidth]{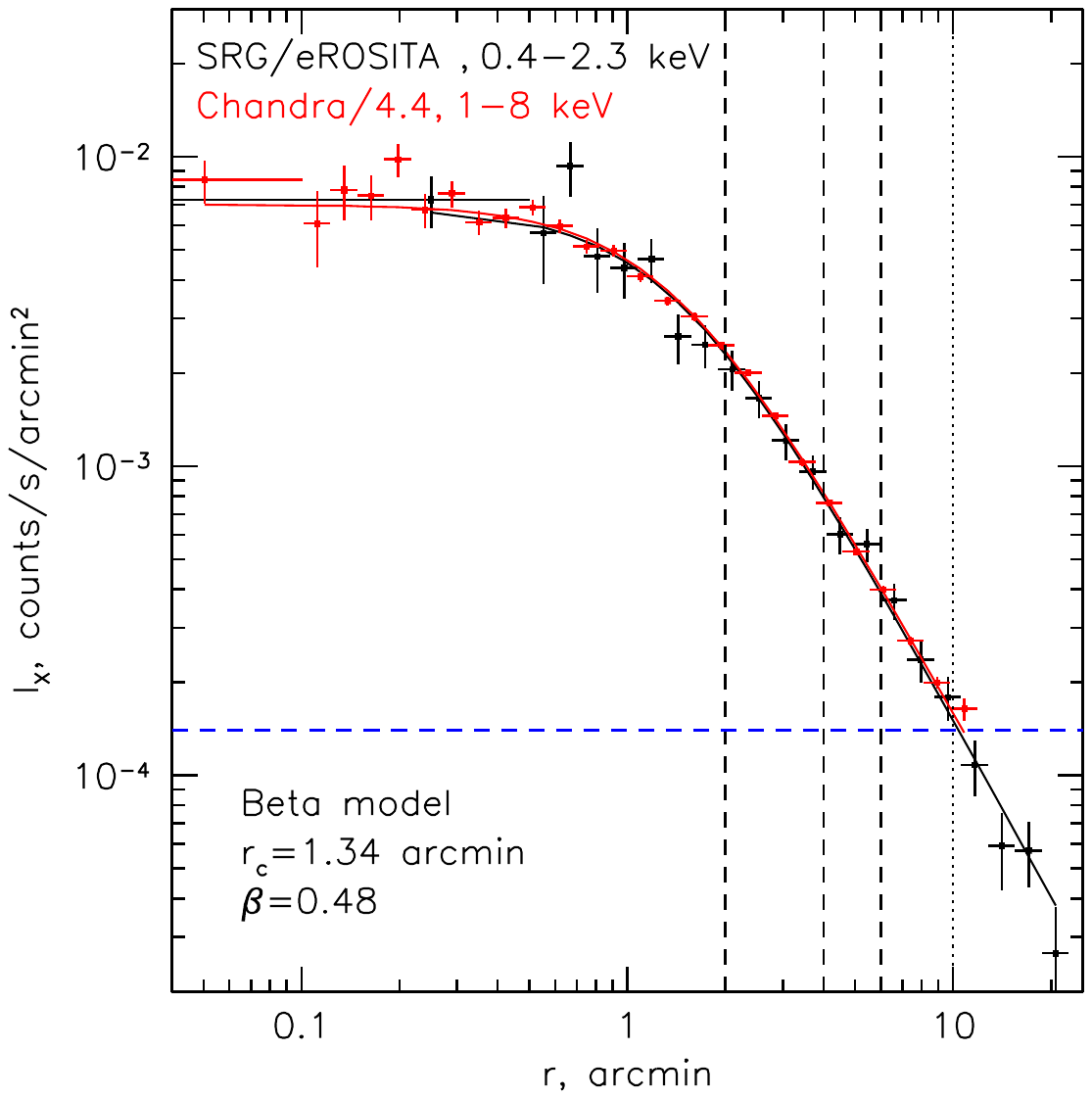}
\includegraphics[angle=0,,clip=true,trim=2.3cm 7.5cm 2.9cm 6cm,width=0.9\columnwidth]{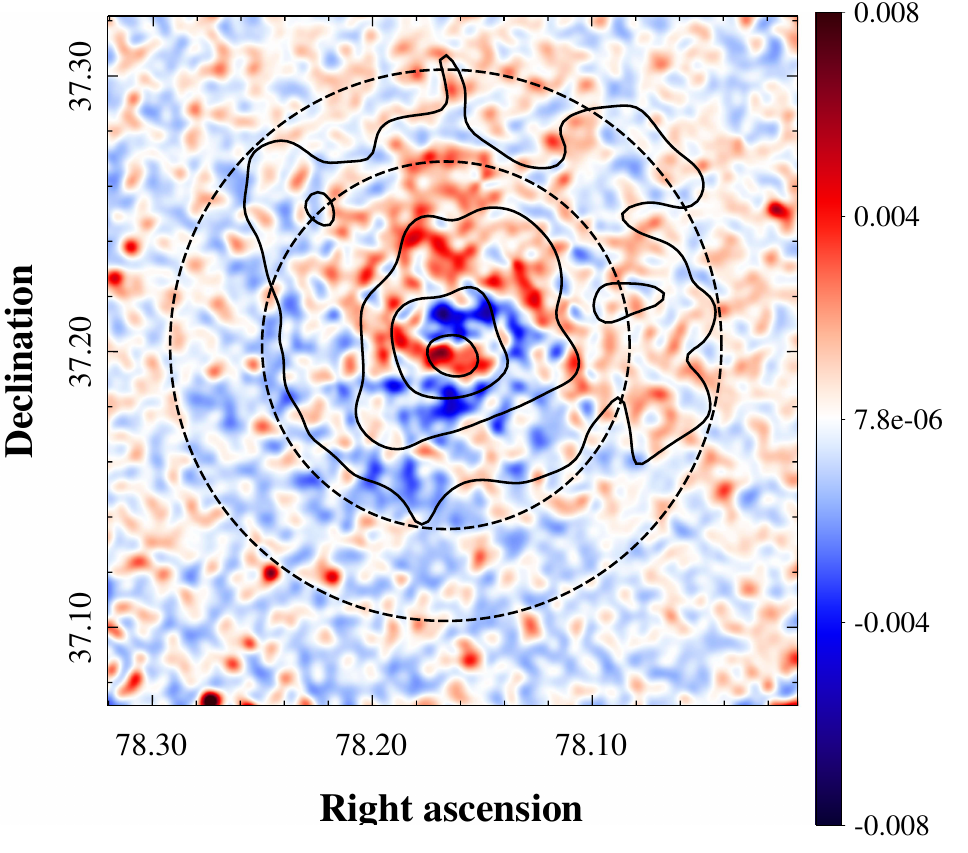} 
    \caption{Summary of the follow-up \textit{Chandra} imaging data. \textit{Top.} Background-subtracted vignetting-corrected image of X-ray surface brightness in 1-8 keV smoothed with a $\sigma=8''$ gaussian. The white circles are $4'$ and $6'$ in radius, the contours show the distribution of 0.4-2.3 keV X-ray surface brightness from the SRG/eROSITA data (see Fig.~\ref{fig:srge_2MASX}).  \textit{Middle.} Radial profiles from \textit{Chandra} (red) and SRG/eROSITA (black) data, with the \textit{Chandra} profile scaled down by a factor 4.4. The lines show the best-fitting $\beta$ profiles, the blue line shows the background level estimated for the SRG/eROSITA profile at large radii. \textit{Bottom.} The residual \textit{Chandra} image after subtracting the best-fit $\beta$ profile.
    }
    \label{fig:cxcimage}
\end{figure}

Following the discovery of the candidate massive galaxy cluster in ZoA, a short 40 ks exploratory observation with \textit{Chandra} was conducted with the aim point at the candidate BCG galaxy 2MASS~J05123831+3711581 (see Appendix~\ref{a:chandra} for details). Due to the exquisite angular resolution and stable background, \textit{Chandra} excels in mapping the diffuse emission above 1 keV, which is of the highest relevance here due to the high absorbing column density and likely high temperature of the cluster.

\subsubsection{Imaging}

Figure~\ref{fig:cxcimage} shows a 1-8 keV X-ray image obtained by \textit{Chandra} with an overlay of contours of the 0.4-2.3 keV surface brightness from SRG/eROSITA. The overall morphology shows a rather smooth and azimuthally symmetric distribution of X-ray surface brightness. The radial profile (centered on the candidate BCG galaxy) very well described by the standard $\beta$ profile $S_{\rm X}(r)=S_0 (1+(r/r_c)^2)^{-3\beta+1/2}$ \citep[][]{1976A&A....49..137C} with $\beta=0.48$ and $r_c=1.34'$, {corresponding to 115 kpc} (see Fig.~\ref{fig:cxcimage}). The same profile fits the radial profile in the 0.4-2.3 keV band from the SRG/eROSITA data remarkably well, showing that the cluster emission dominates over the background within a radius of $10'$, i.e., it fills the whole field-of-view of \textit{Chandra}. This estimate is likely a lower limit, since the sky background in the direction of the Galactic plane includes a part that is less absorbed than the cluster's emission in the same band (and the cosmic X-ray background as well).

Stacking analysis of the X-ray surface brightness profiles in 0.4-2.3 keV has shown that the point where X-ray surface brightness of a cluster (independent of the distance at low redshifts) is equal to the level of the sky X-ray background is close to its $R_{\rm 500c}$ radius \citep[][]{2023MNRAS.525..898L}. As a result, we can see that for \CL~, $R_{\rm 500c}\approx10'$, corresponding to 860 kpc, which is consistent with the expectation for a massive cluster with $M_{500c}\approx2.5\cdot10^{14}M_\odot$. On the other hand, the core radius $r_c$ of the profile is sensitive to the dynamical state of the cluster \citep[e.g.,][]{1984ApJ...276...38J}, and for \CL~ it is a few times smaller than for Coma \citep[canonical non-cool core cluster with $r_c\approx290$ kpc, e.g. ][]{2021A&A...651A..41C} but several times larger than for Perseus \citep[canonical cool core cluster with $r_c\approx28$ kpc, e.g.][]{2026A&A...707A.381C}, while having a relatively shallow value of $\beta\approx0.5$ (similar to Perseus having $\beta\approx0.51$).     

One can use the measured azimuthally-averaged profile as a model and subtract it from the data to highlight possible substructures and global deviations from it.

The right panel in Fig.~\ref{fig:cxcimage} shows such residuals, which appear reminiscent of the famous central mini-spiral, as well as a global sloshing dipole seen in Perseus \citep[][]{2026A&A...707A.381C}. 

 At the same time, the \textit{Chandra} image shows that the diffuse X-ray emission is not peaked on the BCG galaxy, and it does not host a prominent compact X-ray source, in agreement with the picture of the soft X-ray emission by SRG/eROSITA (also, no point source that could be associated with HD~280580 star is visible by \textit{Chandra}, which might be an indication of that emission being transient, consistent with a stellar flare origin). Thus, we conclude that the dynamical state of \CL~ might be relatively perturbed, probably as a result of the ongoing assembly process, and this conclusion can be tested independently by means of X-ray spectroscopy. 

\subsubsection{Spectroscopy}

We start with spectroscopy of the central $r=2'$ region, which encompasses the core radius of the $\beta$ profile. This is the brightest and quite compact region of the cluster, so the influence of background estimation and instrumental effects is minimized here. The emission from this region is well described by an absorbed single-temperature collisional ionization equilibrium (CIE) model \texttt{APEC} \citep[][]{2001ApJ...556L..91S} in XSPEC \citep[][]{1996ASPC..101...17A} in the form of \texttt{wabs(N$_{\rm H}$)*apec(kT,A,z,S)}. The best fit parameters of the model are total absorbing column density $N_{\rm H}=(8.4\pm0.08)\cdot10^{21}$~cm$^{-2}$, gas temperature $kT=(5.3\pm 0.3)$ keV, abundance of heavy elements $A=0.4\pm0.1$ with respect to the Solar abundance \citep[][]{1989GeCoA..53..197A}, redshift $z=0.067\pm0.007$, and absorption-corrected average surface brightness $S_{0.5-2}=10^{-12.44\pm0.01}$~erg~s$^{-1}$~cm$^{-2}$~arcmin$^{-2}$ in 0.5-2 keV ($S_{2-8}=10^{-12.74\pm0.03}$~erg~s$^{-1}$~cm$^{-2}$~arcmin$^{-2}$ for 2-8 keV).
The obtained parameters are virtually independent of the assumptions made regarding the background emission, which include pure particle-induced background (PIB) only, PIB and expectation for the Cosmic X-ray background (CXB), or estimated directly from an annulus of $r=6'-8'$. \ik{Contribution of the Galatic Ridge emission in the direction of $l\approx170^{\circ}$ is expected to be at least a factor of 10 sub-dominant, based on the well established X-ray-IR correlation \citep[][]{2006A&A...452..169R}.}

First, this measurement provides an independent determination of the cluster's redshift, which turns out to be fully consistent with the optical spectroscopic measurement within the uncertainties, which are still quite big for the X-ray measurement (driven mostly by the location of the Fe~XXV line, sensitive to proper calibration of the detector's temperature-dependent spectral properties). Thus, we can confirm the association of the X-ray emitting gas with the candidate BCG and red-sequence galaxies.

Second, the measured temperature is quite high, corresponding to a cluster with the mass of $M_{500c}\approx4.2\cdot10^{14}M_{\odot}~(kT_X/{\rm 5~keV})^{3/2}=4.6\cdot10^{14}M_{\odot}$ according to the standard scaling relations \citep[e.g.,][]{2006ApJ...640..691V}, with an estimated scatter of order of 20\% \citep[][]{2006ApJ...650..128K}. Given that no peaking X-ray emission is observed, this temperature estimate is not expected to be strongly biased down due to the presence of a cool core. For such a cluster, one expects $R_{500c}\approx1.1$ Mpc, corresponding to $r\approx 12.8'$ at $z=0.0745$, very close to the estimate obtained from the radial profile. The ratio of $R_{500c}/r_c\approx9.4$ is indeed intermediate to those found for Coma and Perseus clusters \citep[e.g.,][]{1984ApJ...276...38J,2021A&A...651A..41C,2026A&A...707A.381C}.

\begin{figure}
    \centering
    \includegraphics[angle=0,,clip=true,trim=0.5cm 5.2cm 1.2cm 3.5cm,width=1.\columnwidth]{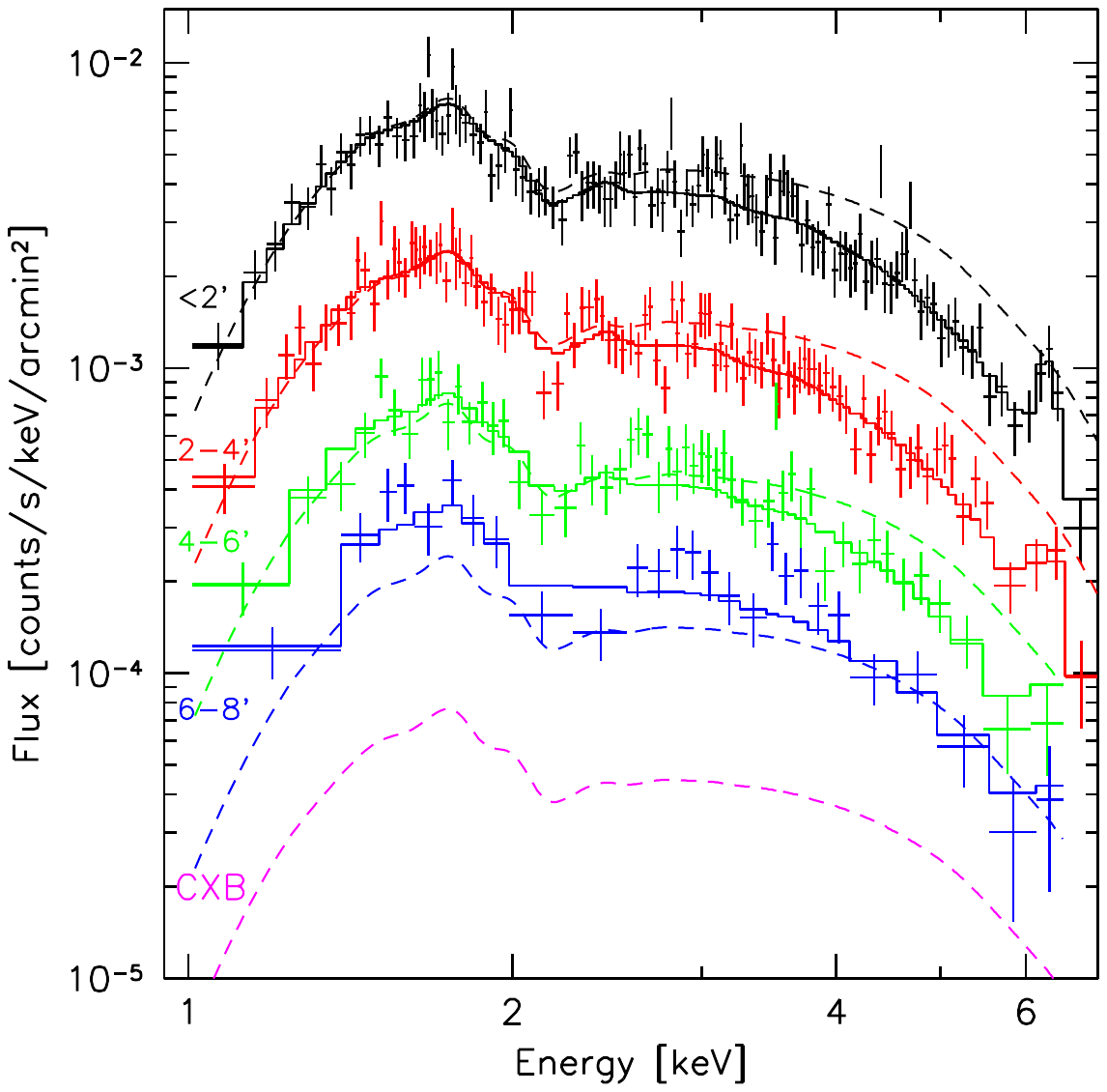}

    \caption{X-ray spectroscopy of the cluster's emission. The data points correspond to the rings of  $0'-2'$ (black), $2'-4'$ (red), $4'-6'$ (green), and $6'-8'$ (blue)  in radius after subtraction of the expected particle background. The solid histograms show the best-fit single temperature models. Magenta dashed curve shows expected level of CXB \citep[e.g.,][]{2006ApJ...645...95H} with the same absorption, while the other dashed curves show CXB scaled up by factors of 10$^{0.5n}$, with $n=1-4$, from down to up, respectively, for relative guidance.
    }
    \label{fig:cxcspectra}
\end{figure}

We proceed with the spectroscopy of X-ray emission in a number of rings with boundaries at r=2, 4, 6, and 8$'$ (see Fig.~\ref{fig:cxcspectra}). Since the data quality and ratio to the background decrease significantly at outer radii, we fix all the parameters of the single-temperature \texttt{APEC} fits except for the temperature and normalization. Also, a number of point sources are masked down to the SRG/eROSITA sensitivity of $2\cdot10^{-14}$~erg~s$^{-1}$~cm$^{-2}$, which corresponds to $\sim30\%$ of the CXB being resolved \citep[e.g.][]{2003ApJ...588..696M}, supressing its expected contribution with respect to the total flux of $1.7\cdot10^{-11}$~erg~s$^{-1}$~cm$^{-2}$~deg$^{-2}$ in the 2-8 keV band least affected by Galactic absorption \citep[][]{2006ApJ...645...95H}. For relatively small \ik{solid angles} considered here, though this effect is likely smaller due to low number statistics of expected sources in these apertures \citep[e.g., ][]{2025A&A...702A.175L}   

While the evolution of the normalization and the corresponding average surface brightness is fully consistent with the radial profile, no strong temperature evolution is observed with $kT_{2-4}=(4.3\pm0.4)$ keV, $kT_{4-6}=(4.3\pm0.9)$ keV, $kT_{6-8}=(4.4\pm2.4)$ keV for r=$2'-4'$, $4'-6'$, $6'-8'$ rings, respectively. For all of them, the cluster's emission is higher than the expected CXB level by at least a factor of several, and should be well within the $R_{500c}$ radius of the cluster. Thus, there is no indication of the core being cooler than the bulk of the cluster, and, taking at face value, it is even slightly hotter, which might be an indication of a recent shock heating event. \ik{The quality of the available data does not allow us measuring radial profile of the gas metallicity, which can be used to reveal metallicity excess in the center indicative of a relaxed state of the core \citep[e.g., ][]{2001ApJ...551..153D}.}

A convenient property of the $\beta$ profile with $\beta=0.5$ is that it is readily integrated analytically to obtain total flux within a given aperture:
\begin{equation}
    F_{\rm X}(r)=\int_0^{r}2\pi r S_{\rm X}(r)=\pi r_c^2 \int_0^{\xi_c^2}\frac{d\xi^2}{1+\xi^2}=\pi r_c^2 S_0\ln(1+\xi_c^2),
\end{equation}
with $\xi_c=r/r_c$, which is logarithmically-divergent, implying that some steeping of the profile should happen at large radii. Spectral modeling for the innermost region of the cluster in the manner similar to the one described above allows one to estimate $S_0$ very robustly and with a few percent accuracy, giving estimates of $S_{0,\,0.5-2}=3.4\cdot10^{-13}$ erg~s$^{-1}$~cm$^{-2}$~arcmin$^{-2}$ and $S_{0,\, 2-8}=4.8\cdot10^{-13}$ erg~s$^{-1}$~cm$^{-2}$~arcmin$^{-2}$ in the 0.5-2 keV and 2-8 keV bands, respectively. Integration within a radius which is 10 times bigger than $r_c$ gives total fluxes of $F_{0.5-2}=8.9\cdot10^{-12}$ erg~s$^{-1}$~cm$^{-2}$ and $F_{2-8}=1.1\cdot10^{-11}$ erg~s$^{-1}$~cm$^{-2}$ in the 0.5-2 keV and 2-8 keV bands. At $z=0.0745$, these fluxes correspond to luminosities $L_{0.5-2}=1.2\cdot10^{44}$ erg~s$^{-1}$ and $L_{2-8}=1.6\cdot10^{44}$ erg~s$^{-1}$. These values are in perfect agreement with expectations for a $M_{500c}=(4.5\pm0.5)\cdot10^{14}M_{\odot}$ cluster from the standard scaling relations, which are known to have larger scatter of $\sim50\%$ for a fixed mass in this case though \citep[e.g.,][]{2006ApJ...640..691V}.

\section{Discussion}
\label{s:discussion}

\begin{figure}
    \centering
    \includegraphics[angle=0,,clip=true,trim=2.5cm 6.8cm 2.1cm 4.cm,width=0.99\columnwidth]{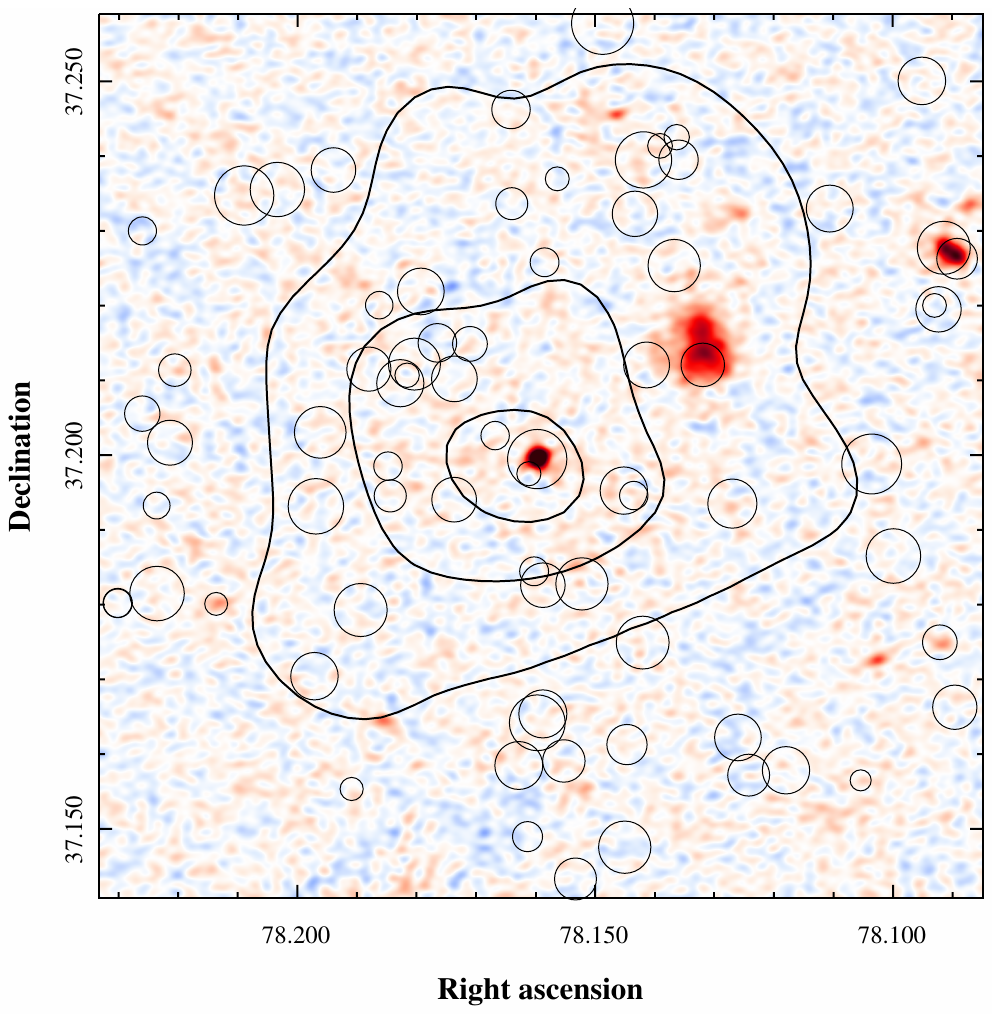}      
    \caption{Radio picture of the central part of \CL~ with the LoTSS 150 MHz radio map as background. The circular regions and contours are the same as in Fig.\ref{fig:ukidss_srgecont_2masx}.  In addition to the brightest source in the center coincident with the candidate BCG, a number of fainter sources might be associated with candidate red-sequence galaxies. A large diffuse radio source is visible to the north-west of the cluster center (see Sec.~\ref{ss:uss}) and might be associated with a candidate red-sequence galaxy as well (although with a substantial offset).
    }
    \label{fig:lotss_srgecont_2masx}
\end{figure}

\begin{figure}
    \centering
    \includegraphics[angle=0,,clip=true,trim=1.5cm 6.8cm 2.5cm 5.cm,width=0.99\columnwidth]{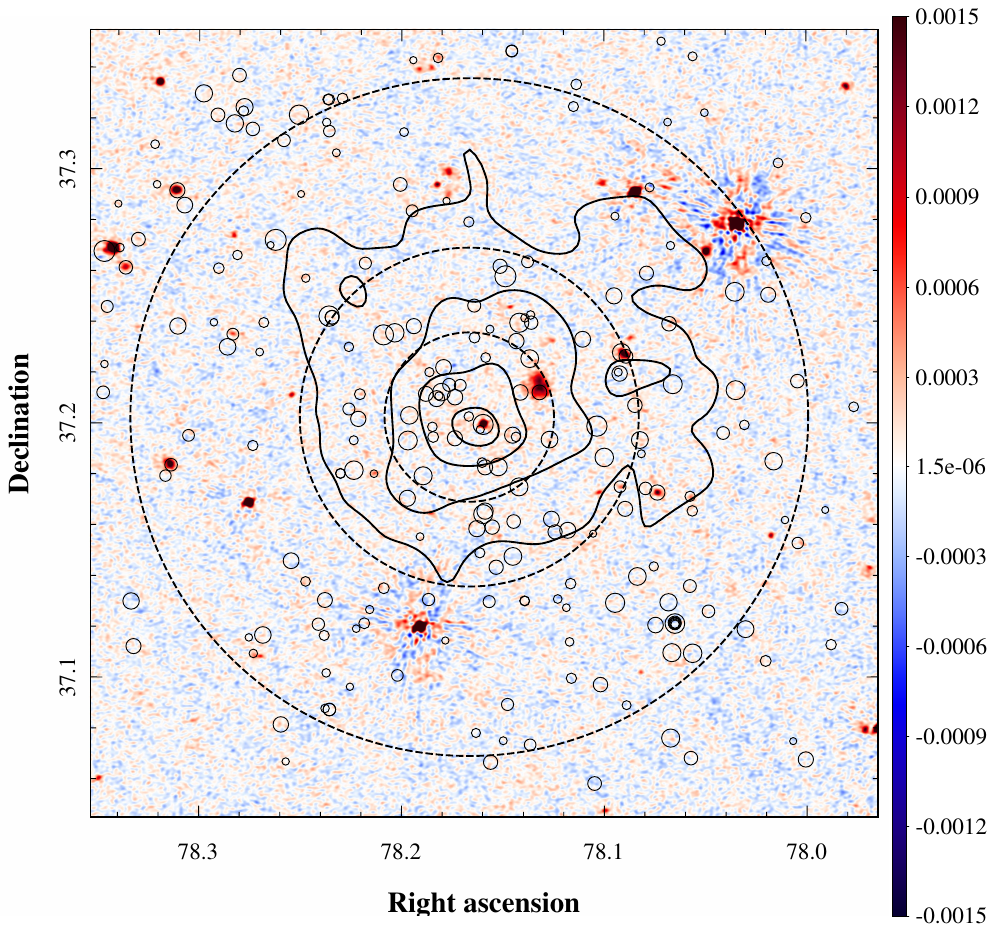}
    \caption{LoTSS 150 MHz image at larger scales. Concentric circles are 2$'$, 4$'$, 8$'$ in radius. Small circles are UKIDSS red sequence galaxies. Solid black contours show levels of X-ray surface brightness.}
    \label{fig:lotss_large}
\end{figure}

\subsection{Summary of the properties}
\label{ss:summary}

The follow-up optical and X-ray observations allowed us not only to confirm the nature of the newly found object as a massive galaxy cluster, but also to derive its properties in a very reliable manner. These properties can be summarized as \CL~ is likely a massive non-cool core cluster with $M_{500c}=(4.5\pm0.5)\cdot10^{14}M_{\odot}$ and central temperature of $kT=5.3\pm0.5$ keV at redshift $z=0.0745$. With the 0.5-8 keV unabsorbed flux at the level of $F_{0.5-8}=2\cdot10^{-11}$ erg~s$^{-1}$~cm$^{-2}$ and corresponding luminosity of  $L_{0.5-8}=2.8\cdot10^{44}$ erg~s$^{-1}$, this cluster would have been among several dozens of the brightest clusters on the sky \citep[e.g.][]{1984ApJ...276...38J,1998ApJ...503...77M}. An example of \CL~ shows that even such a massive cluster could be missing as a result of high absorption and confusion in the ZoA. 

Another efficient method of identification of massive galaxy clusters is via the thermal Sunyaev-Zeldovich effect \citep[][]{1972CoASP...4..173S}, which, however, also suffers from extremely bright and structured diffuse emission in the Galactic plane \citep[][]{2016A&A...594A..22P,2023MNRAS.526.5682C}. Interestingly, there is a positive signal in the direction of \CL~ in the maps of the $y$ parameter obtained by component separation analysis of the \textit{Planck} maps (see Appendix~\ref{s:planck}). The amplitude of this signal is quite high, with the peak value of $y_{0}\approx 1.6\times10^{-5}$, and not far from expectations based on the scaling relations \citep[e.g., ][]{2021ApJS..253....3H} taking into smearing with the PSF of $Planck$ ($FWHM\approx10'$). \ik{We can also use the X-ray surface brightness profile and spectrally measured temperature to infer the central gas density, which turns out to be relatively low, $n_{e,0}\approx0.004$~cm$^{-3}$, corresponding to the central value of the $y$ parameter $y_0\approx5\cdot10^{-5}$ within the core radius of $r_c\approx1.35'$. For comparison, the central value of the $y$ parameter for Coma (which is fully resolved by $Planck$) is at the comparable level of $y_0\approx6\cdot10^{-5}$ \citep[][]{2013A&A...554A.140P}, while the scaling with the mass is $\propto M_{500c}^{4/3}$.}  However, the presence of large-scale variations of positive and negative amplitude due to contaminating foregrounds makes the statistical significance of possible detection quite low. Future observations with higher angular resolution might be instrumental in confirming and possibly separating and mapping this signal in more detail \citep[][]{2019SSRv..215...17M}.        

\subsection{Candidate Diffuse Radio Source with Steep Spectrum }
\label{ss:uss}

Many massive clusters are actively assembling in the current epoch, sometimes in a very violent manner. As a result, a large amount of kinetic energy of the infalling structures is converted into heat, but also in turbulence and relativistic particles. Some of the CIZA clusters are especially known for possessing radio-emitting sub-structures \citep[e.g.,][]{2010Sci...330..347V}, which partially comes from the fact, that identification of a cluster in the ZoA is strongly eased when a characteristic radio features are observed (in other words, there should be many more clusters which are unspectacular in radio but they are just unknown).   

Given that optical/NIR and X-ray properties of \CL~ suggest that it might not be in a fully settled configuration, one can ask if there are any prominent radio substructures visible in it, especially at low frequencies powered by relatively long-living relativistic electrons. Examination of the LoTSS~DR3 image at 150 MHz \citep[][]{2026arXiv260215949S} indeed shows a presence of a diffuse radio source (decomposed in two sources, ILT~J051231.57+371252.5 and 	ILTJ051232.97+371245.6 , in the LoTSS DR3 catalog) at the projected distance of $\approx2'=170$ kpc from the cluster's center (see Fig.~\ref{fig:lotss_srgecont_2masx}). No indication of a similar structure at higher frequencies, e.g., in the RACS maps at 887.5 and 1367.5 MHz (see  Fig.~\ref{fig:srge_ps1}), is visible, suggesting a rather steep slope of the radio emission. The total extent of the diffuse source is $\sim1'$, corresponding to $\sim90$ kpc, and a couple of candidate cluster member galaxies are located next to it. Thus, this source might be an example of stripped lobes inflated before the infall into the cluster or a very old and risen bubble, comparable in terms of radial distance and size to the one in the Ophiuchus cluster \citep[][]{2020ApJ...891....1G}. \ik{Given the large variety of diffuse radio sources with steep spectra revealed recently in LOFAR observations of galaxy clusters \citep[e.g.][]{2016MNRAS.459..277S,2025MNRAS.538.3326W}, the sparsity of the currently available data for \CL~ prevents us from drawing firm conclusions about the origin of this source.}   

On larger scales (see Fig.~\ref{fig:lotss_large}), one can spot a number of candidate red-sequence galaxies coincident with tail-like radio sources, which might be examples of narrow-angle tail radio galaxies. Global shock-like structures are not visible, although interferometric filtering of $\sim10'$ scales might also hinder their detection. If the possible shock is not far from the cluster's center, or if it propagates along the line of sight, it can be rather problematic to see characteristic signatures in radio emission. Deeper X-ray observations, especially equipped with high-resolution spectroscopic capabilities, would be instrumental to check this.

\subsection{Location in the local large-scale structure}
\label{ss:lss}

Massive galaxy clusters are not only interesting objects themselves, but also important signposts of the large-scale structures of the Universe. They are located in the nodes of the cosmic web and sometimes form even larger conglomerates, but not necessarily gravitationally bound, known as superclusters \citep[e.g.][]{2025A&A...695A..59B}.  

The 2MASS Redshift Survey  \citep[2MRS, ][]{2012ApJS..199...26H} was particularly instrumental in reconstructing LSS in the local Universe \citep[e.g.,][]{2015AJ....149..171T}, but it does not cover the ZoA region as well. In this regard, it is interesting to locate CIZA clusters identified by other independent methods within the 2MRS-reconstructed web and compare their locations with probable expectations \citep[see e.g. an example of 3C129 cluster in ][]{2020A&A...644A.107R}.

\begin{figure*}
    \centering
    \includegraphics[width=1.0\linewidth]{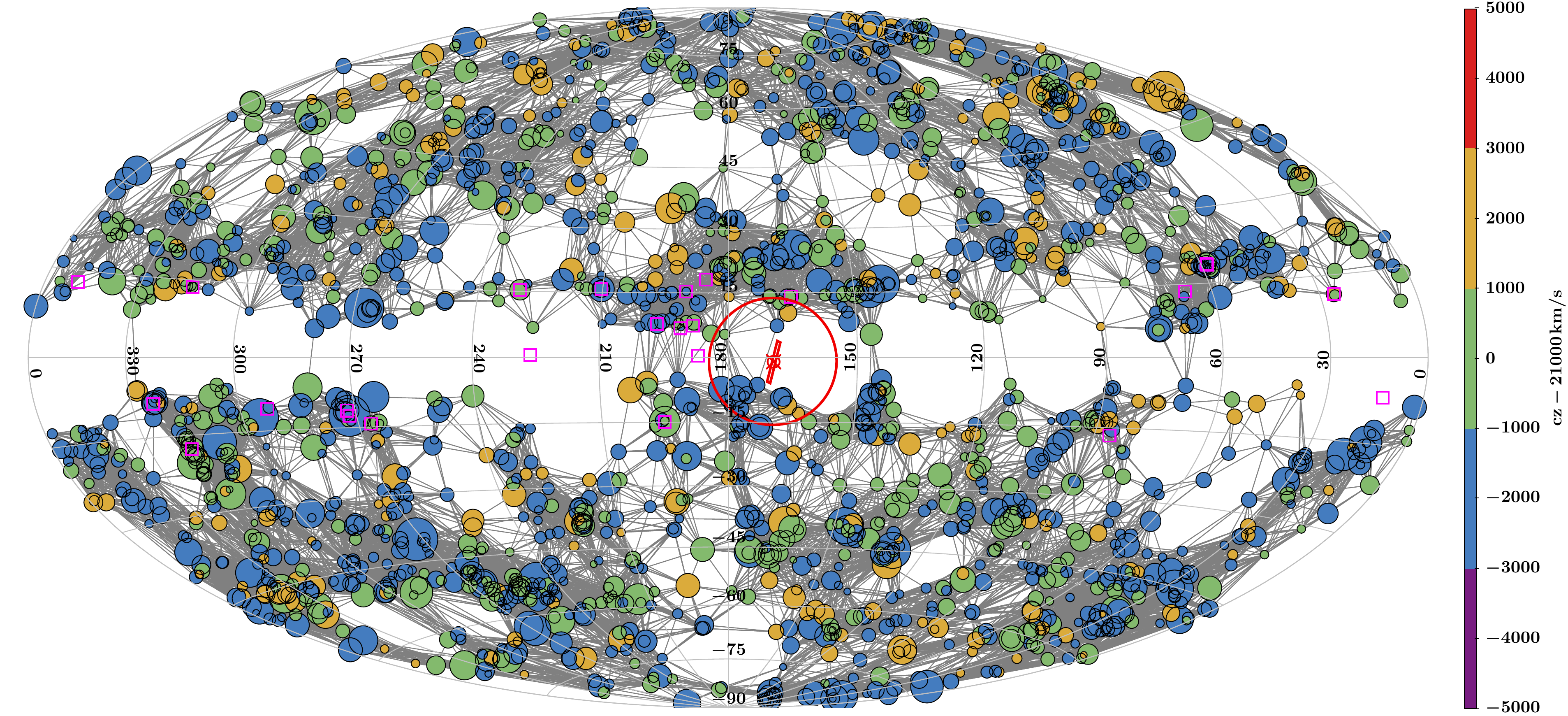}

    \caption{All-sky map (Galactic coordinates, centered on $(l,b)=(180^\circ,0^\circ)$, i.e. on the Galactic anti-center) of galaxies from 2MASS Redshift Survey \citep[][]{2012ApJS..199...26H} with the redshift in the range from 0.06 to 0.08. All pairs of galaxies closer to each other that 15 deg  ($\approx75$ Mpc) are connected by a gray line to highligh LSS's connectivity. The red cross marks the position of \CL~ with the bar showing the direction along  $PA=40^\circ$ (corresponding to elongation in distribution of candidate cluster galaxies),  while the big circle has a radius of 15$^\circ$, i.e., 75 Mpc at this redshift. Color-coding reflects LOS velocity difference with respect to z=0.07, and the size of the points is a proxy for their $K$ band magnitude. Magenta boxes mark positions of known CIZA clusters in the same redshift bin \citep[][]{2002ApJ...580..774E,2007ApJ...662..224K}.}
    \label{fig:2mrs_ciza}
\end{figure*}

In order to do so, we consider all 2MRS galaxies within a 0.06-0.08 redshift bin, and find their connectivity with the tophat window with radius $r=15$ deg scale, corresponding to 75 Mpc at $z=0.07$, which is also sufficient to connect galaxies on different sides of the $|b|<5^\circ$ ZoA in the 2MRS data \citep[][]{2012ApJS..199...26H}.  Figure~\ref{fig:2mrs_ciza} shows all these galaxies (in Galactic coordinates with the center at $(l,b)=(180^\circ,0^\circ)$, i.e., at the Galactic Anti-Center) with color-coding reflecting their residual LOS velocity (with respect to $z=0.07$) and the size being a proxy for the K band luminosity. All pairs of galaxies closer than 15 deg on the sky are connected with a gray line in this figure, and one can easily see connected over-dense or isolated under-dense regions. The lines connecting objects across the Galactic plane might be considered as likely "routes" for finding spines of the LSS in these regions. In other words, given the current information in the maps, these locations appear to be the most plausible candidates for finding overdensities of galaxies, galaxy clusters, and groups. On the other hand, given that the bias factor of the massive clusters with respect to the general galaxy distribution is high, it is quite unlikely that they will be found in places other than these.

This picture is confirmed by the already known CIZA clusters  \citep[][]{2002ApJ...580..774E,2007ApJ...662..224K}  shown in pink in Fig.~\ref{fig:2mrs_ciza}, which is, of course, partially an observational effect since the same 2MRS data were partially used to facilitate discovery for some of them.  It is stunning that \CL~ is indeed not only located in one of the high probability locations, but also that the elongation axis of the cluster's member galaxies (shown in red in Fig.~\ref{fig:2mrs_ciza}) perfectly matches the expectation from the LSS on the much larger physical scale. 

\CL~ appears to be a dominant node of its LSS environment, as highlighted by the red circle in Fig.~\ref{fig:2mrs_ciza} marking a radius of 75 Mpc, which corresponds to half of the characteristic LSS wavelength. On a slightly large scale, it might be a part of even larger, $\gtrsim$100 Mpc, structure that contains a number of galaxy clusters at the redshift between 0.06 and 0.08 projecting towards the Galactic Anti-Center. Given that this region of the sky is not strongly affected by Galactic emission and absorption, exploration of this region by future soft X-ray microcalorimetric missions with high grasp \citep[e.g., Line Emission Mapper,][]{2022arXiv221109827K,2024SPIE13093E..27K} might be particularly promising for the detection of warm-hot intergalactic medium \citep[][]{2023arXiv231016038K}. Indeed, the redshift window around $z=0.07$ is one of the optimal ones for the detection of O~VII and O~VIII against much stronger Galactic foreground emission \citep[][]{2024SPIE13093E..27K}.  

Finally, this picture predicts a number of likely locations of other CIZA candidates in the Galactic plane, e.g., at $l=330^\circ, 280^\circ, 200^\circ, 145^\circ, 115^\circ, 90^\circ $ or $ 70^\circ$. The application of the technique presented here (combining data of shallow X-ray, radio, and NIR/optical observations) should be able to check this prediction in a straightforward manner. The resulting augmented picture can be compared with the constrained cosmological simulations of the Local Universe \citep[e.g.][]{2023A&A...677A.169D}, for many of which structures at redshifts beyond 0.07 are located close to the boundary of the simulation's domain and are not strongly fixed due to sparsity of the constraints in the observational data. In particular, a comparison of the LSS in such simulations at the $\sim$100 Mpc scale \citep[e.g., ][]{2025A&A...702A.243S} with the possible supercluster-like structure in the Galactic Anti-Center direction, as well as the most probable locations of massive clusters behind the Galactic plane, might provide important insights and possible improvements. 

\ik{A similar approach can be applied to look for candidate CIZA clusters at lower or higher redshifts. In the same time, the overall volume of the ZoA is smaller at lower redshifts, meaning that only a handful of massive clusters are expected to be found there, with a number of them already well known, e.g. the 3C~129 cluster. On the other hand, for $z>0.1$, the width of the ZoA becomes too large on the physical scale, what should diminish correlation between structures on opposite sides of the ZoA and consequently the predictive power of this simple connectivity technique.}

\section{Conclusions}
\label{s:conclusion}

We report the discovery of a massive galaxy cluster in the Galactic ZoA using the data of the SRG/eROSITA all-sky survey. The BCG of the cluster was identified among the extended sources in the 2MASS and UKIDSS surveys based on the presence of radio emission in the RACS survey. Optical spectroscopy of the candidate BCG provided determination of the redshift of the cluster $z=0.0745\pm0.001$. 

Follow-up observations with the \textit{Chandra} X-ray observatory confirmed the SRG/eROSITA findings, while also offering an opportunity to measure X-ray temperature, luminosity, and size ($R_{500c}$) of the cluster, which are all consistent with its mass being at the level of ${M_{500c}=\rm(4-5)\times10^{14}~M_{\odot}}$, while also probably indicating an unrelaxed dynamical state. The inferred relatively low central gas density is yet another indicator of the unrelaxed state.

The location of this cluster and the elongation of its galaxy population (selected via NIR red sequnce on the magnitude-color diagram) are consistent with an expectation from the large-scale structure at this redshift, and it might be part of an extended overdensity of such objects in the Galactic Anticenter direction. Examination of X-ray, radio, and infrared data in the locations of the ZoA, where similar objects are expected to be found based on the LSS properties, might reveal another $\sim10$ clusters at this redshift in the future.

Deeper follow-up observations in optical, X-ray, radio, and sub-mm domains will be needed to clarify the nature of the number of identified features (e.g., substructures in the X-ray image and the candidate ultra-steep diffuse radio source), while it can also be used as a bright and extended background source with a relatively simple spectrum for the Galactic absorption studies.

\begin{acknowledgements}


In this work, observations with the eROSITA telescope onboard \textit{SRG} space observatory were used. The \textit{SRG} observatory was built by Roskosmos in the interests of the Russian Academy of Sciences represented by its Space Research Institute (IKI) in the framework of the Russian Federal Space Program, with the participation of the Deutsches Zentrum für Luft- und Raumfahrt (DLR). The eROSITA X-ray telescope was built by a consortium of German Institutes led by MPE, and supported by DLR. The SRG spacecraft was designed, built, launched, and operated by the Lavochkin Association and its subcontractors. The science data are downlinked via the Deep Space Network Antennae in Bear Lakes, Ussurijsk, and Baikonur, funded by Roskosmos. The eROSITA data used in this work were converted to calibrated event lists using the eSASS software system developed by the German eROSITA Consortium and analysed using proprietary data reduction software developed by the Russian eROSITA Consortium. \\

The authors are grateful to TUBITAK, IKI, KFU, and the Tatarstan Academy of Sciences for partial support in the use of RTT-150 (Russian-Turkish 1.5-m telescope in Antalya). 
The work of IB, EI, MS was supported by subsidy  30000P.16.1.OH17AA81027, allocated to TAS IAS to fulfill the state assignment in the field of scientific activity. \\

Observations at the SAO RAS telescopes are supported by the Russian Ministry of Science and Higher Education. The instrumentation upgrade is being implemented as part of the national project "Science and Universities."

IK was supported by the Simons Foundation via the Simons Investigator Award to A. A. Schekochihin.

WF and RK acknowledge support from the Smithsonian Institution, the \textit{Chandra} High Resolution Camera Project through NASA contract NAS8-0306, NASA Grant 80NSSC19K0116 and \textit{Chandra} Grant GO1-22132X. \\

This work is based on publicly available optical data from the Pan-STARRS Surveys.
The Pan-STARRS1 Surveys (PS1) and the PS1 public science archive have been made possible through contributions by the Institute for Astronomy, the University of Hawaii, the Pan-STARRS Project Office, the Max-Planck Society and its participating institutes, the Max Planck Institute for Astronomy, Heidelberg and the Max Planck Institute for Extraterrestrial Physics, Garching, The Johns Hopkins University, Durham University, the University of Edinburgh, the Queen's University Belfast, the Harvard-Smithsonian Center for Astrophysics, the Las Cumbres Observatory Global Telescope Network Incorporated, the National Central University of Taiwan, the Space Telescope Science Institute, the National Aeronautics and Space Administration under Grant No. NNX08AR22G issued through the Planetary Science Division of the NASA Science Mission Directorate, the National Science Foundation Grant No. AST-1238877, the University of Maryland, Eotvos Lorand University (ELTE), the Los Alamos National Laboratory, and the Gordon and Betty Moore Foundation.
\end{acknowledgements}

\bibliographystyle{aa}
\bibliography{current} 

\begin{appendix}
\section{SRG/eROSITA all-sky survey data }

We use the data from the eROSITA telescope \citep{2021A&A...647A...1P} on board the  \textit{SRG} mission  \citep{2021A&A...656A.132S}, launched in 2019 which started to perform the all-sky survey mission in December 2019. We use the data accumulated over four consecutive scans, with the total effective exposure amounting to $\approx 850$ seconds per point.  Initial reduction and processing of the data were performed using standard routines of the \texttt{eSASS} software \citep{2018SPIE10699E..5GB,2021A&A...647A...1P}, while the imaging and spectral analysis were carried out with the background models, vignetting, point spread function (PSF) and spectral response function calibrations built upon the standard ones via slight modifications motivated by results of calibration and performance verification observations \citep[e.g.][]{2021A&A...651A..41C,2023MNRAS.521.5536K}.
\section{Follow-up observations with Chandra}
\label{a:chandra}

\CL~ was observed by the Chandra Advanced CCD Imaging Spectrometer (ACIS) four times with varying roll angle but with the same aim-point at 2MASS ~05123831+3711581 (see Table~\ref{tab:chandra_obsid} for details). Observations were processed with the standard Chandra data reduction {(
CIAO v. 4.17) and calibration software (CalDB v. 4.12). }
Data analysis steps are described in detail in  \cite{2009ApJ...692.1033V}  and include high background period filtering, application of the latest calibration corrections to the detected X-ray photons, and determination of the background intensity in each observation.  For spectral analysis, we generated the spectral response files that combine the position-dependent ACIS calibration with the weights proportional to the observed brightness. The total filtered exposure time is $\approx 39$~ks (as summarized in Table~ \ref{tab:chandra_obsid}). 

\begin{table}
\centering
\renewcommand{\arraystretch}{1.1}
\caption{Details of the Chandra observations of \CL. }
\begin{tabular}{cccc}
\hline
ObsID & Instrument & Mode & Exposure, ks        \\ \hline
30507 & ACIS-I & FAINT & 9.7 \\
30666 & ACIS-I & FAINT & 9.0 \\
30667 & ACIS-I & FAINT &  10.1   \\
30668 &	ACIS-I & FAINT &	9.2 \\
\hline 
\end{tabular}
\label{tab:chandra_obsid}
\end{table}

%
\section{Optical observations}
\label{s:optical}

The RTT-150 observations were carried out on November  01, 2024, and January 18, 19, 20, 2025, with the TFOSC instrument and the Andor iKon-L 936 BEX2-DD-9ZQ CCD with a size of $2048\times 2048$ pixels,  thermoelectrically cooled to $-$80$^{\circ}$C. The field of view, in direct image mode, is $11'\times11'$ with the scale of 0.33 arcsec/pixel at binning 1$\times$1.
For each of the four observed galaxies (see Sec.~\ref{ss:opticalspectra}, Fig.~\ref{fig:ukidss},  2MASX~J05124037+3712199, candidate BCG 2MASS~J05123831+3711581,
2MASS~J05122483+3711555 and 2MASS~J05125014+3714048), three spectra with a single exposure of 3600 sec were obtained by using grism 15 and a 134$\mu$ entrance slit (corresponding to 2.4 arcsec at the sky). The wavelength range is 3900 - 8900 \r{A}, and the spectral resolution is 15 \r{A} (binning 2$\times$2 was used in spectral observations). Spectrophotometric standard star BD17d4708 was observed at RTT-150 each night for flux calibrations. Spectroscopic data were reduced using standard tools of the IRAF package\footnote{The IRAF astronomical data processing and analysis package is available at \url{https://iraf-community.github.io}}. Observed fluxes were corrected for extinction $E(B-V) = 1.1$. The strongest absorption lines (H-alpha, H-beta, G-band, MgB, NaD, Fe I, Mg I, and Ca I) were identified in the galaxy spectra, and their measured wavelengths were used for redshift measurements. The accuracy of the redshift measurements is 0.001.

\begin{figure}
  \centering
    \includegraphics[width=1\columnwidth]{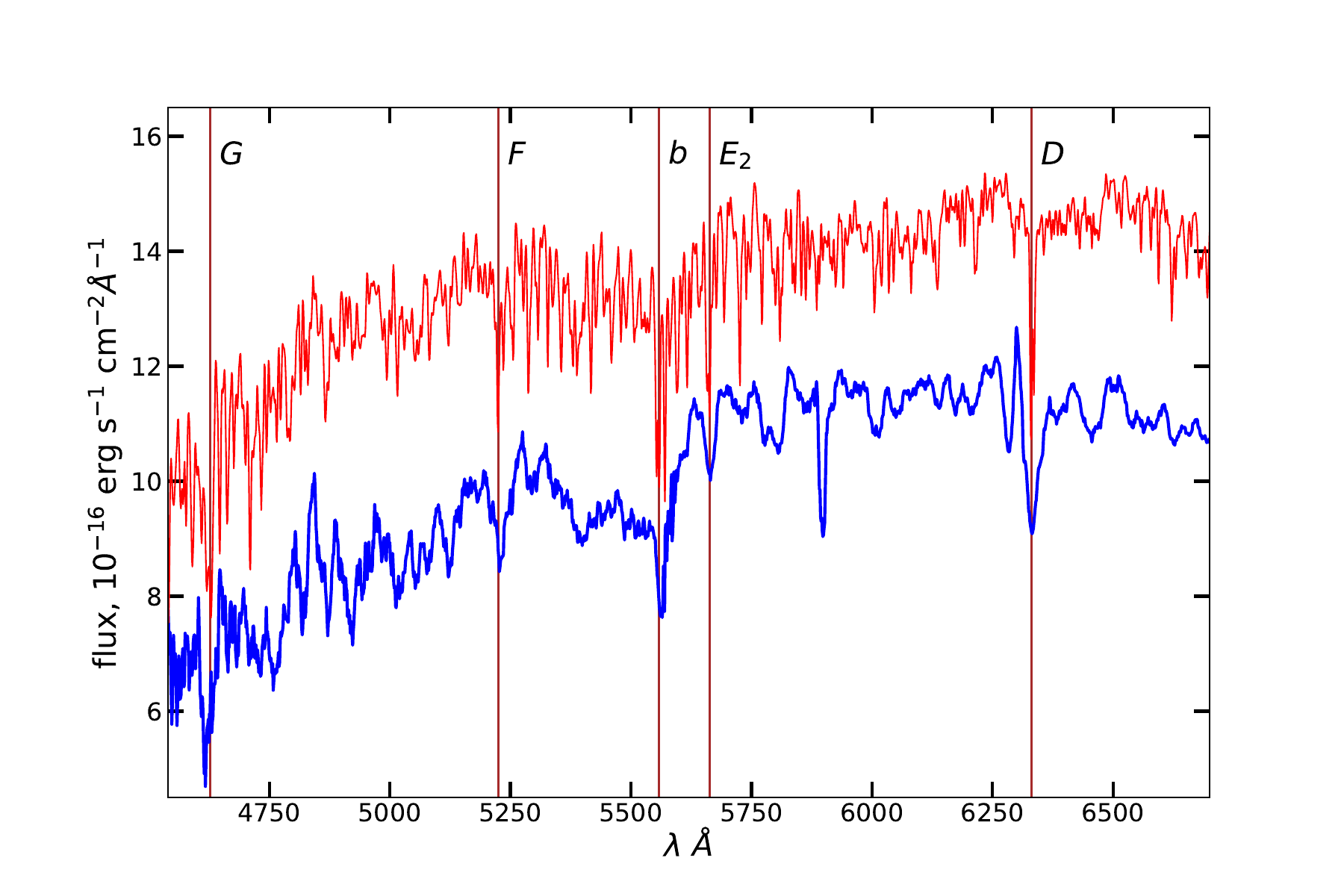}
    \caption{Optical spectrum of the candidate BCG obtained with the BTA telescope corrected for interstellar extinction (blue) in comparison to the spectrum of a synthetic stellar population (red) with metallicity $Z=0.02$ aged by 11 billion years \citep{2003MNRAS.344.1000B}, redshifted to $z = 0.07$. The vertical lines indicate the Fraunhofer lines.
    }
  \label{fig:spec_BCG}
\end{figure}

The BTA observations were carried out on 26 January 2025 with the SCORPIO-2 spectrograph \citep{2011BaltA..20..363A} at the 6-m BTA. {The detector was E2V  CCD261-84\citep{2023Photo..10..774A}}. We used the VPHG940@600 grism and {a $1\farcs5$ width} long slit. The wavelength coverage was 3500--8500~\r{A}, with a spectral resolution of 10.4~\r{A}. The slit was positioned at $\alpha = 00{:}00{:}32.0$, $\delta = +00{:}00{:}57.0$ (ICRS, J2000.0) with a position angle of $89.1^{\circ}$ covering the candidate BCG.
Eight exposures of 900~s each were obtained. The spectrophotometric standard star G191B2B was observed the same night for flux calibration. Data reduction was performed using IRAF and custom software. All fluxes were corrected for Galactic extinction using $E(B-V)=1.1$. The redshift of the BCG, $z=0.0745\pm0.0010$,  was also determined by identifying the strongest absorption lines (shown in the Fig.~\ref{fig:spec_BCG}).

\section{Possible Sunyaev-Zeldovich signal}
\label{s:planck}
Although Sunyaev-Zeldovich (SZ) effect \citep[][]{1972CoASP...4..173S} has proven to be extremely efficient in detecting massive galaxy clusters, the very high level of foreground Galactic emission at mm and sub-mm wavelengths makes blind detections nearly impossible in the ZoA. Here, being guided by X-ray, NIR, and optical observations, we perform a check for possible traces of the thermal SZ signal in the direction of the component-separated $y$-map by $Planck$ \citep[PR4, ][]{2023MNRAS.526.5682C}.

\begin{figure}
    \centering
    \includegraphics[trim=2.5cm 7.cm 4cm 5cm,width=0.95\columnwidth]{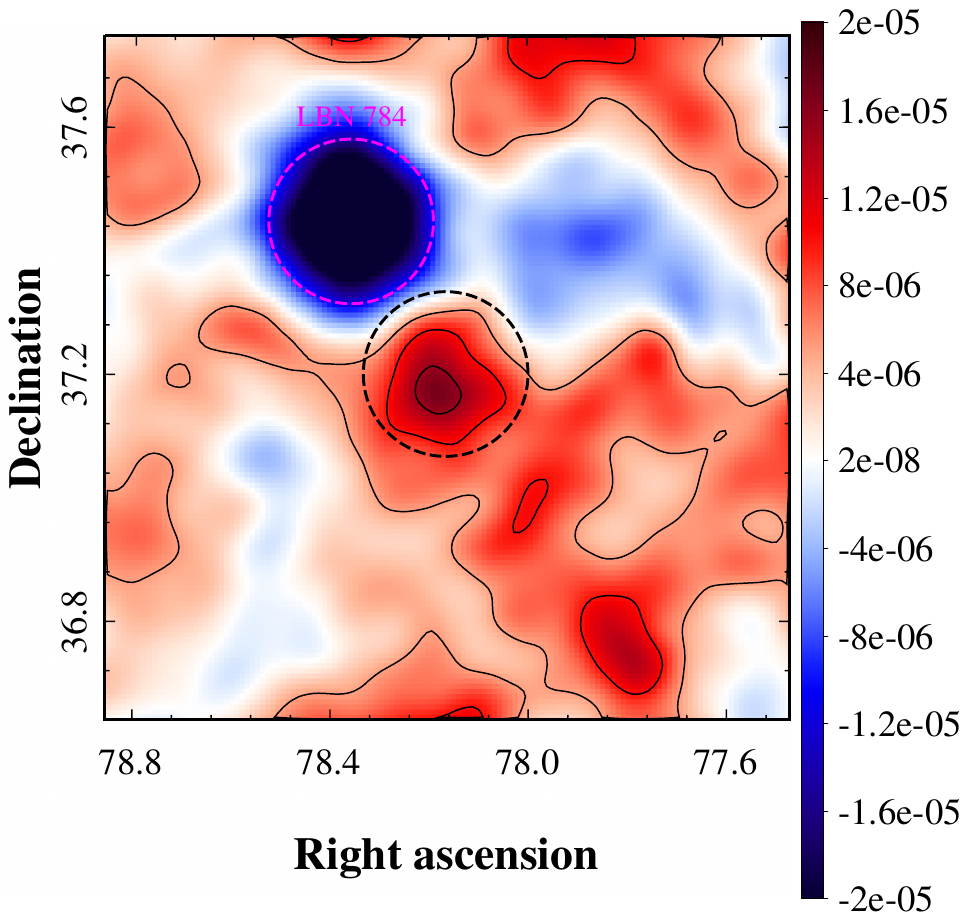}
    \caption{A map of the Sunyaev-Zeldovich $y$-parameter based on $Planck$ component-separated all-sky data \citep[PR4, ][]{2023MNRAS.526.5682C}. The contours are at levels of $0.5,1,$ and $1.5\times10^{-5}$. The dashed black circle is centered on  CIZA~J0512.7+371 and has a radius of 8', while the magenta circle (of the same radius) highlights the location of a foreground HII region LBN~587.}
    \label{fig:planck}
\end{figure}

Figure~\ref{fig:planck} shows that there is a marginal indication of positive excess in the direction of \CL, although it barely exceeds the rms level of fluctuations at a similar scale (taking into account that the effective resolution of the $Planck$ map does not allow resolving structures smaller than several arcmin in size). Additional complexity comes from bright foreground "negative" sources, arising due to imperfections of the component separation procedures, as exemplified by the bright HII region LBN 784 \citep[Sharpless~228, e.g. ][]{1990ApJ...362..147C}, which is a prominent source of thermal (dust) and non-thermal (synchrotron) emission.

Although any robust interpretation of this signal is quite challenging, future dedicated high-resolution multi-frequency observations might be capable of confirming or disproving it, especially when guided by the X-ray, NIR, and radio data.

\end{appendix}

\end{document}